\documentclass[man,12pt,floatsintext]{apa7}
\usepackage{ragged2e}
\usepackage{setspace}  

\makeatletter
\AtBeginDocument{%
  \ifx\processdelayedfloats\@undefined
  \else
    \renewcommand\processdelayedfloats{}
  \fi
}
\makeatother

\usepackage[american]{babel}
\usepackage{csquotes} 
\usepackage[style=apa,sortcites=true,backend=biber,natbib=true]{biblatex}
\usepackage{geometry}
\usepackage{array}
\usepackage{booktabs}

\usepackage[T1]{fontenc} 
\usepackage{mathptmx} 
\usepackage{longtable} 
\usepackage{booktabs} 

\DeclareUnicodeCharacter{0301}{}

\hypersetup{hidelinks}

\title{Organizational Structure and the Research of U.S. Library and Information Science Schools} 
\shorttitle{Organizational Structure of LIS Schools and LIS Research} 
\author{Jiangen He$^{1}$, Wen Lou$^{2}$}
\affiliation{$^{1}$School of Information Sciences, {The University of Tennessee, Knoxville}, {TN, USA}\\
$^{2}$School of Information and Communication, Nankai University, {Tianjin, China}}

\abstract{U.S. Library and Information Science (LIS) schools are organized differently: some are independent, and others share a parent unit with computing, communication, education, or the arts and sciences. We ask whether their research differs accordingly, using 18,048 publications (2013--2026) by 1,602 faculty members at 44 ALA-accredited schools. We grouped the 15,934 publications with abstracts into 15 research themes with document embeddings and hierarchical clustering, and classified venues by field. Faculty members' research profiles differ by school type in themes and, more clearly, in venues, but type accounts for only 2.2\% and 5.3\% of the variation; most lies among faculty of the same school. The differences run from libraries and archives to computational research. Schools housed with computing spread over the most themes and publish most on security, networks, and urban computing, on clinical and biomedical informatics, and in computer science venues; schools housed with education spread over the fewest and concentrate most on libraries, archives, and computing education; independent schools stay closest to the average. Over the period, schools moved away from library research and LIS venues, but seldom toward the venues of their parent unit's discipline. Research appears to follow a school's faculty more than its parent unit.}

\begin{document}

\maketitle 
\justifying

\section{Introduction}

Library and Information Science (LIS) is among the most self-examined fields in the social sciences. Scholars have debated what information is and whether LIS should be understood as a science, a profession, or a meta-discipline \citep{Borko1968,Saracevic1999,Bates1999,Furner2015}. They have also mapped the intellectual terrain of the field through bibliometric analysis, co-word analysis, topic modeling, and content analysis \citep{Larivire2012,Milojevi2011,Sugimoto2011,han2020evolution,jarvelin2022lis,Wu2025}. These studies have documented changes in research topics and methods and have provided different ways of describing the boundaries and internal structure of LIS.

At the same time, the institutions that house LIS have changed. Free-standing library schools have been merged into larger academic units, absorbed by other departments, or reorganized as iSchools \citep{Hildreth2002Organizational,Cronin2005,Goulding2018Location,wang2025ischools}. In the United States, LIS faculty now work in units with different disciplinary homes, including standalone schools and units located within Arts \& Sciences, Education, Communication, and Computing. This organizational diversity creates a question that is less visible in field-level studies: whether the institutional setting in which LIS research is produced is associated with the research portfolio of the school.

The literature on LIS research and organization have largely developed in parallel. Research landscape studies describe what LIS researchers study, but their units of analysis are usually the field, a country, a journal set, or a body of publications. The institutional units employing those researchers therefore receive limited attention. Studies of LIS organization, meanwhile, have examined restructuring, curricula, accreditation, governance, professional education, and disciplinary identity \citep{Hildreth2002Organizational,Goulding2018Location,Cronin2005,pryor2022reaffirming}. They provide evidence about where LIS is located and how its institutional identity is interpreted, while giving less attention to how organizational location relates to research.

There are reasons to examine the two dimensions together. The sociology of science has long treated intellectual and social organization as connected features of scientific fields \citep{whitley2000intellectual}. Organization theory similarly describes organizational structure as a context that shapes information processing, attention, and decision-making \citep{Joseph2020Organizational,Joseph2025}. LIS research provides some evidence consistent with this perspective. \citet{astrom2008formalizing}, for example, found a connection between academic affiliation and research orientation in Nordic LIS. Studies of LIS faculty have also documented differences in disciplinary backgrounds and academic positioning across units \citep{Wiggins2012,Zhu2016,zuo2019standing}. More recently, \citet{he2025academic} provided a preliminary comparison of academic structure and research focus in U.S. LIS schools. These studies point toward a relationship between institutional setting and research, but they do not establish how that relationship appears across the U.S. LIS field.

In academic settings, organization is where faculties are affiliated, study at, and publish with. In the meantime, faculties are the main composition of what organizations stand for. Organizations' disciplinary context can derive from disciplinary composition and scholarly positioning of faculties \citep{Sengupta2017Choice,Kienast2023}. We classify 44 U.S. LIS units into five organizational types and analyze 18{,}048 publications of 2013--2026 by 1{,}602 of their faculty members. This study examines whether the research portfolios of the units differ across these organizational structures, whether the venues in which their faculty publish differ in the same way, and whether units of different types changed differently over the period. We treat this relationship at the institutional-unit level and faculty as the research-producing unit. Publications provide evidence of individual faculty research, individual publication profiles are aggregated to the unit level, and the resulting distribution of topics represents a unit's research portfolio. This study connects three strands of evidence that have usually been examined separately: the intellectual landscape of LIS, the institutional organization of LIS education, and the organizational context of knowledge production.

\section{Literature Review}

\subsection{LIS disciplinary background and research profiles}

LIS has no single settled intellectual definition.
Some accounts place information itself at the center of the field \citep{Borko1968,Saracevic1999,Furner2015}, while others emphasize the organization, use, and communication of information \citep{Bates1999}.
Others stress the field's ties to neighboring disciplines and its borrowing from the social sciences \citep{Cronin2008}, or its development through continuing interaction among information science, librarianship, and professional practice \citep{Bawden2008}.
This pluralism matters because institutional units may differ in disciplinary orientation even when they share the same field label.

Empirical studies have repeatedly found internal structure within the LIS research landscape.
\citet{Milojevi2011} identified library-, information-, and bibliometric-oriented branches through title-word clustering.
\citet{Sugimoto2011} identified changing research areas in North American LIS dissertations, and \citet{han2020evolution} traced topic changes in LIS research from 1996 to 2019 with latent Dirichlet allocation.
\citet{jarvelin2022lis} examined 50 years of LIS journal articles through content analysis, and \citet{Wu2025} described the field through three dimensions centered on library, people, and algorithm.
Although these studies differ in corpus, period, geographic scope, and method, each offers a structure for understanding the field.

The resulting themes span established areas of LIS and newer forms of information research.
Information literacy is a longstanding area of research on people's engagement with information \citep{Martzoukou2017}, and human--information interaction is another established direction \citep{Marchionini2008}.
More recent work has addressed online communities, computational methods, artificial intelligence, and other emerging information environments \citep{chen2024you,tomavsev2020ai,Xu2025}.
National and regional studies further show that these areas receive different levels of attention across contexts \citep{Dora2020,Mirmani2024,Rehman2024,Kassim2024}.

The methods used to construct these landscapes have also changed.
Earlier studies relied on manual content analysis, bibliometric indicators, word frequencies, and latent topic models, whereas recent studies use document embeddings and neural topic models to group semantically related publications at scale \citep{Jung2024,Babalola2024,Kim2024,Kimura2024,Xu2025}.
Comparisons of topic-modeling approaches show that model choice and parameter settings can affect topic coherence and interpretation \citep{Borin2024,Azhar2025,Yang2025}.
A different set of themes may therefore reflect differences in data, method, or level of analysis as well as in the underlying research.

Three observations from this literature are relevant here.
First, LIS research has expanded toward computational and human-centered topics alongside established areas of librarianship and information research \citep{han2020evolution,ma2021evolution}.
Second, disciplinary background is associated with the topics and methods represented in LIS research, raising questions about fragmentation within the field \citep{Vakkari2022Disciplinary,vakkari2023association,vakkari2024characterizes}.
Third, research profiles vary across national and regional settings \citep{Dora2020,Mirmani2024,Rehman2024,Kassim2024}.
These findings leave open whether the organizational structure of a unit within a university is also associated with this variation.

\subsection{Institutional organization of LIS schools}

The institutional organization of LIS has received sustained attention, particularly in relation to the movement from independent library schools toward larger academic units. \citet{Hildreth2002Organizational} documented organizational realignment from standalone units to incorporated programs, while \citet{Goulding2018Location} examined how organizational location affects LIS programs. The emergence of the iSchool movement generated further discussion of disciplinary identity and institutional positioning \citep{Cronin2005,Dillon2012}, including comparisons of schools' research, education, and organizational missions \citep{wang2025ischools}. Hybrid academic units that cross established disciplinary boundaries have also been examined \citep{pryor2022reaffirming}.

Much of this literature focuses on education and institutional identity, examining the development of data-oriented programs, archival education, doctoral training, professional preparation, accreditation, and governance \citep{li2022worldwide,wiles2024teaching,Partridge2024Documenting,salaba202321st,Lieutenant2016}. This emphasis reflects the direct relationship between organizational location and educational responsibilities. It also suggests that organizational structure may shape faculty hiring through curricular and programmatic needs. Under this interpretation, differences in research portfolios may emerge partly through the disciplinary composition of faculty recruited to meet those needs.

Research on faculty composition provides a bridge between organizational structure and research. Studies of LIS faculty hiring and academic backgrounds show changes in the disciplinary composition of the field and differences among institutional settings \citep{Zhu2016,Wiggins2012,zuo2019standing}. \citet{Song2025} also examined factors associated with research productivity among LIS faculty in North American universities. At the publication level, \citet{walters2016disciplinary} showed that departments make uneven contributions to LIS literature. These studies show that faculty composition and institutional units matter for LIS scholarship, but they do not connect organizational type with the topical portfolio of a school. \citet{he2025academic} provides an initial step in this direction by examining academic structure and research focus. The present study extends this line of inquiry by applying a common organizational classification and faculty-level aggregation across U.S. LIS units.

\subsection{Organizational structure and knowledge production}

Research outside LIS provides a broader basis for examining organizational structure as a context for knowledge production. Organization theory connects organizational structure with information processing and decision-making \citep{Joseph2020Organizational,Joseph2025}. Structures vary in how they configure authority, control, coordination, and channels of information \citep{Wagner2021Dimensional,Joseph2025}, and because they change in response to institutional pressures, an observed structure also reflects earlier organizational change \citep{Luciano2026Why}.

Studies of research organizations provide more direct evidence of connections between organizational arrangements and scholarly activity.
Laboratory-level research has examined relationships between research organization and academic production \citep{carayol2004does,verbree2015organizational,braam2014indicators}.
Other studies have related administrative practices, institutional change, authority, and funding arrangements to research performance \citep{Amara2015What,onder2008institutional,jimenez2024organizational,Cruz-Castro2018Autonomy,glaser2018changing}.
Research on universities likewise points to organizational culture and management as contexts for academic research \citep{Kienast2023}, and studies of research centers, knowledge-transfer offices, and research portfolios take a similar organizational view \citep{TorresZapata2019,Sengupta2017Choice,Ekboir2017Knowing}.

This literature suggests that organizational structure may be associated with research, but its evidence has limits for LIS.
It concerns research productivity, performance, or collaboration more than research topics, and its units of analysis are laboratories, research centers, or universities rather than academic units whose faculty come from several disciplinary traditions.
Moreover, organizational structure and faculty selection are difficult to separate empirically: faculty backgrounds may shape the character of a unit, while the disciplinary home of the unit may influence whom it recruits.

These considerations point to a gap at the intersection of the LIS and organizational literatures.
LIS landscape studies show that research topics vary across time, disciplinary backgrounds, and geographic settings; studies of LIS organizational change document the movement of programs across academic structures; and research on organizations links institutional arrangements to scholarly activity.
This study connects these observations empirically.

\subsection{Conceptual framework and research questions}

We use \emph{organizational structure} as the overarching concept and classify LIS units into five organizational types according to their disciplinary home within the university: \textsc{Information} (independent units), \textsc{Art\&Science}, \textsc{Education}, \textsc{Communication}, and \textsc{Computer}. We define a school's \emph{research portfolio} as the distribution of its faculty's research across topical themes.
The central relationship is between organizational structure and research portfolio. We conduct this association analysis because organizational restructuring, hiring practices, faculty research interests, and institutional history may influence one another, and a single causal direction would require longitudinal data from all schools in history.
Beyond the research portfolio, we examine the \emph{scholarly positioning} of the units. Faculty in different organizational settings may publish in different venues and address different scholarly communities. Publication venues provide an observable indicator of this positioning and can be compared with the topical portfolios of the units \citep{walters2016disciplinary,Vakkari2022Disciplinary}. Both may be linked to organizational type through \emph{faculty composition}: organizational types may recruit faculty with different disciplinary backgrounds, and disciplinary background has been associated with differences in LIS research topics and methods \citep{Zhu2016,Wiggins2012,zuo2019standing,vakkari2023association}. We do not observe the backgrounds of the faculty, so composition enters the study as an interpretation of the results rather than as a measure.

The framework also recognizes factors that may shape the observed relationship. Institutional research intensity, unit size, hiring needs, and organizational change can influence both faculty composition and publication activity. Temporal changes in a school's portfolio may reflect several processes, including shifts in research interests, faculty turnover, external developments in the field, and organizational restructuring. We do not model these factors; we describe the research intensity and size of the units of each type (Table~\ref{tab:stats}) and consider all of them when interpreting the differences across organizational types and over time.

The research questions follow from this framework:

\begin{enumerate}
\item[\textbf{RQ1.}] To what extent do the research portfolios of U.S. LIS schools differ across organizational structures, both in the themes they emphasize and in how widely their research spreads across themes?

\item[\textbf{RQ2.}] Do faculty in different organizational structures publish in venues of different disciplinary fields, and do these differences correspond to the differences in research themes?

\item[\textbf{RQ3.}] How did the research portfolios and publication venues of U.S. LIS schools change from 2013 to 2026, and did these changes differ across organizational structures?
\end{enumerate}

\section{Methods}

\begin{figure}[!htbp]
    \centering
    \includegraphics[width=\linewidth]{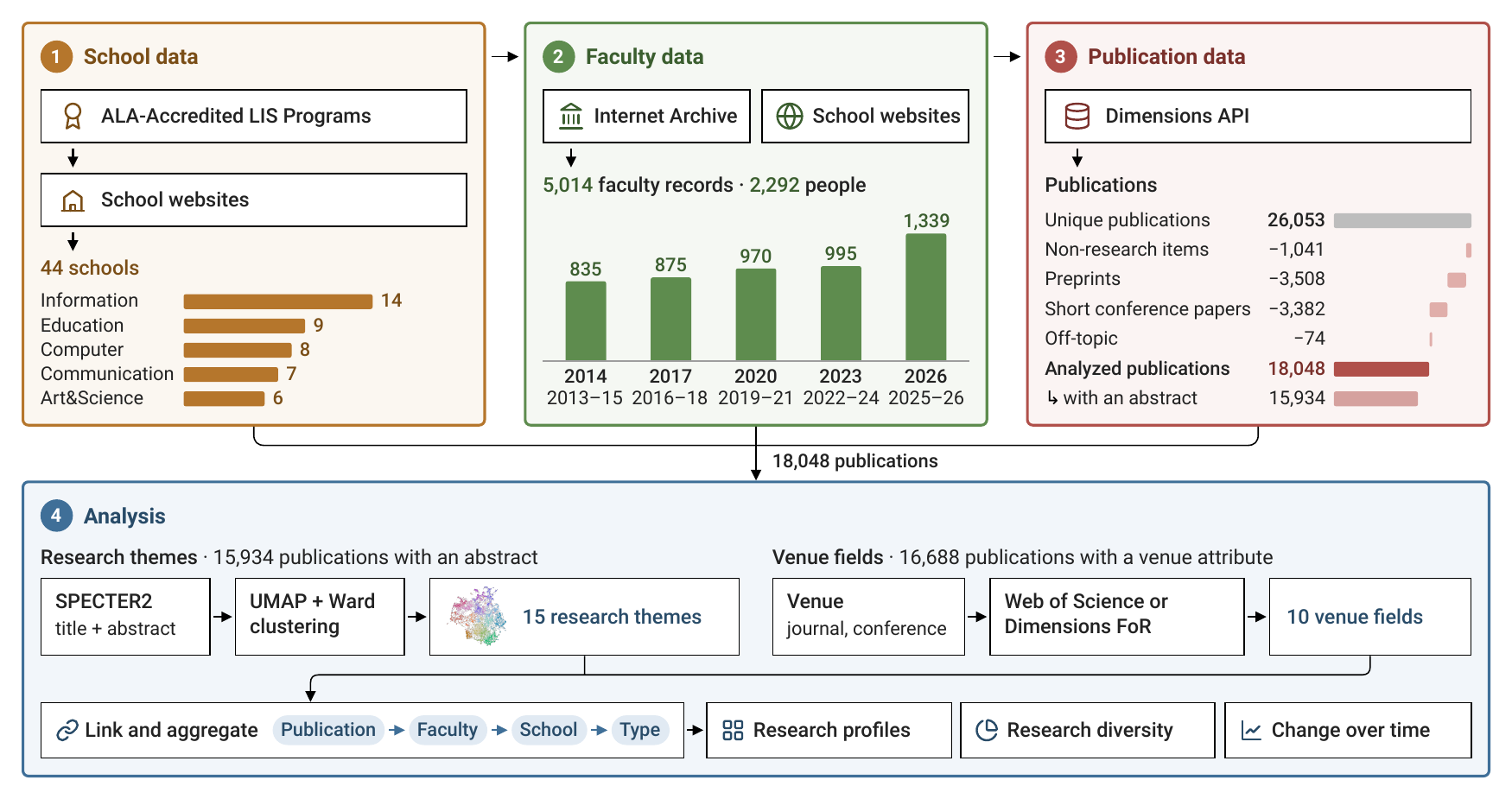}
    \caption{A workflow of the study, including the data collection and analysis.}
    \label{fig:framework}
\end{figure}

The study proceeds in four steps (Figure~\ref{fig:framework}): collecting the LIS schools, their faculty lists over the years, the publications of their faculty from 2013 to 2026, and the analysis.

\subsection{School Data Collection}
We chose the United States as a single case because academic culture, and with it the way universities organize LIS schools, differs greatly from country to country; adding other countries would make the country another major factor in the analysis and would greatly raise the cost of collecting the data.
We started from the 55 U.S. schools on the 2021 list of programs accredited by the American Library Association (ALA) and coded their organizational structure from their current and archived websites (Step 1 in Figure~\ref{fig:framework}).
We excluded 11 schools: three that offer an LIS degree program without an LIS academic unit, most of them without full-time LIS faculty; three without a school website or online faculty information; one that could not be found in web archives (Step 2); and four whose faculty have no publication data in Dimensions (Step 3).
The study thus covers 44 schools.

We developed the organizational types while coding the schools rather than defining them in advance.
For each school, we traced the administrative parent of its LIS unit and the other units under that parent from January 2013 to September 2026, using university announcements, catalogs, and archived websites.
Comparing these records across schools, we grouped the schools by the affiliation arrangements that recurred among them.
Some LIS units report directly to the university, and we labeled this arrangement \textsc{Information}.
Others share a parent with a unit of another discipline: computer science or another computing unit (\textsc{Computer}), communication or journalism (\textsc{Communication}), or education (\textsc{Education}).
A further group sits in a college of arts and sciences or in another college named for a broad area of study, such as the social sciences; because such colleges span too many disciplines for any one sibling unit to indicate a partner discipline, we labeled this arrangement \textsc{Art\&Science}.
Only one other arrangement appeared, a college shared with business, and it applies to a single school, Dominican University, from July 2025, too few to form a type.
Because 17 of the 44 schools changed their parent unit at least once, we coded each period separately and assigned each school the type it held for most of the window, which gives 14 \textsc{Information}, eight \textsc{Computer}, seven \textsc{Communication}, nine \textsc{Education}, and six \textsc{Art\&Science} schools (Table~\ref{tab:classification}).
Table~B1 in Appendix~B gives, for every school, the units that determined its type and every change of parent unit, with the years under each type.

\begin{table}[!htbp]
    \caption{Classification of LIS Schools by Academic Structure}
    \centering
    \small
    \begin{tabular}{p{0.2\linewidth}p{0.08\linewidth}p{0.65\linewidth}}  
    \toprule
    \textbf{Type} & \textbf{Count} & \textbf{Universities} \\
    \midrule
    \textsc{Information} & 14 & Emporia State University,
    North Carolina Central University,
    Simmons University,
    Syracuse University,
    University of Illinois--Urbana-Champaign,
    University of Iowa,
    University of Maryland--College Park,
    University of Michigan--Ann Arbor,
    University of North Carolina--Chapel Hill,
    University of North Texas,
    University of Texas--Austin,
    University of Washington,
    University of Wisconsin--Milwaukee,
    Wayne State University \\
    \midrule
    \textsc{Computer} & 8 & Drexel University,
    Indiana University--Bloomington,
    Indiana University-Purdue University--Indianapolis,
    San Jose State University,
    University at Albany--SUNY,
    University of Hawaii--Manoa,
    University of Pittsburgh,
    University of Wisconsin--Madison \\
    \midrule
    \textsc{Communication} & 7 & Florida State University,
    Kent State University,
    Rutgers University--New Brunswick,
    University of Alabama,
    University of Kentucky,
    University of South Carolina,
    University of Tennessee--Knoxville \\
    \midrule
    \textsc{Education} & 9 & Long Island University Post,
    Louisiana State University--Baton Rouge,
    Texas Woman's University,
    University at Buffalo--SUNY,
    University of California--Los Angeles,
    University of Denver,
    University of Missouri,
    University of North Carolina at Greensboro,
    University of Southern Mississippi \\
    \midrule
    \textsc{Art\&Science} & 6 & CUNY--Queens College,
    The Catholic University of America,
    Dominican University,
    University of Arizona,
    University of Oklahoma,
    University of South Florida \\
    \bottomrule
    \end{tabular}
    \label{tab:classification}
\end{table}

\subsection{Faculty Data Collection}

Because LIS faculty publish across many fields, their publications cannot be collected comprehensively through database categories or venues alone, so we identified the faculty from the schools' websites and collected their publications by name.
Faculty lists change with hiring and departures, so we took them at five points: for 2014, 2017, 2020, and 2023 from archived snapshots of the faculty directory pages in the Internet Archive's Wayback Machine, and for 2026 from the schools' current websites on September 26, 2026 (Step 2 in Figure~\ref{fig:framework}).
Each list stands for a period of publications, the year before, the year of, and the year after it: 2013--2015, 2016--2018, 2019--2021, and 2022--2024 for the lists of 2014, 2017, 2020, and 2023, and 2025--2026, up to the time of collection, for the list of 2026.
Appendix~A describes how we located the archived directory pages. 

An AI browser agent (browser-use) extracted the faculty from each directory page with a standardized prompt (Appendix~A), and we validated every extracted record manually.
We checked the lists for implausible entries and abrupt year-to-year changes and collected anomalous lists again manually.
Thirty-one of the 176 archived lists (44 schools in four years) were not from the right year, because the only snapshot of the page came from another year or the page repeated a later list; we replaced each with a list from an archived page of the correct year or, failing that, from the university's catalog or bulletin for that academic year.
The five lists hold 5{,}014 faculty records (835 in 2014, 875 in 2017, 970 in 2020, 995 in 2023, and 1{,}339 in 2026).
They include tenured, tenure-track, and full-time non-tenure-track faculty, as well as administrators such as deans and chairs who also hold a faculty appointment, and exclude adjunct and visiting faculty, graduate students, and staff.
Emeritus faculty were included when a school's directory listed them among its faculty: 112 of the 2{,}292 faculty members.

\subsection{Publication Data Collection}

We collected the publications of the faculty from the Dimensions Analytics API (Step 3 in Figure~\ref{fig:framework}), which covers journals and conferences broadly and disambiguates authors.
Merging the faculty records across years by name and organization gave 2{,}292 unique faculty members.
We matched each to a Dimensions researcher profile by name and the GRID identifier of their university, verified every match against the profile's names, affiliations, and publication titles, and added further profiles of the same person where Dimensions had split one researcher into several (Appendix~C).
In all, 1{,}602 of the 2{,}292 faculty members (69.9\%) were matched.

\begin{figure}[!htbp]
    \centering
    \includegraphics[width=\linewidth]{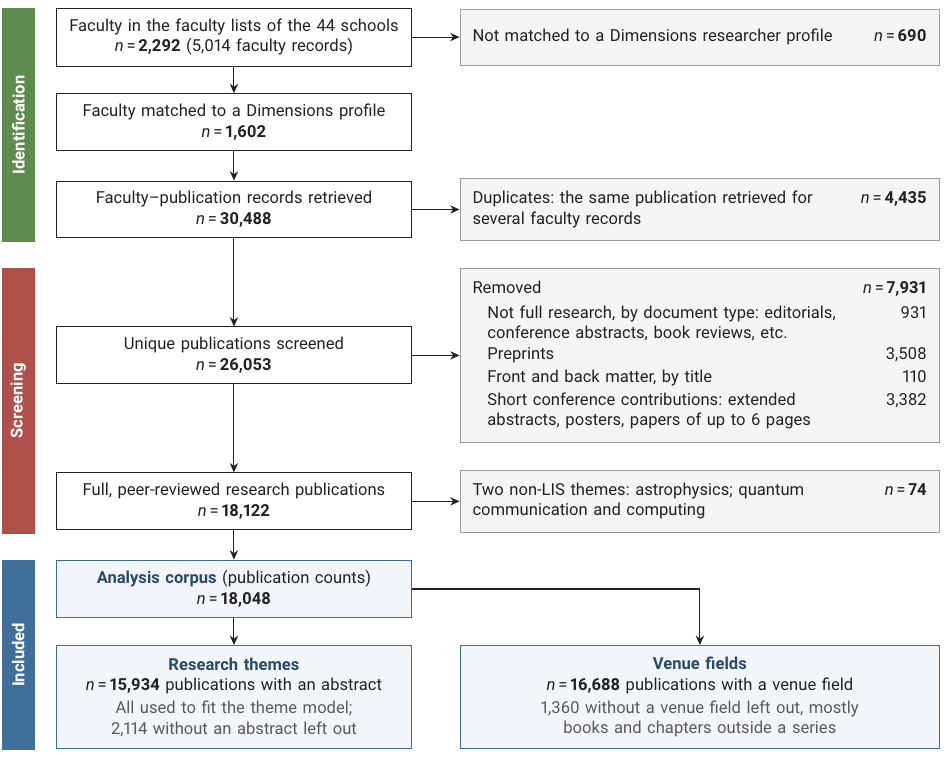}
    \caption{Selection of the publications analyzed, adapted from the PRISMA 2020 flow diagram \parencite{page2021prisma}. Each removed record is counted under the first criterion it meets, in the order listed.}
    \label{fig:pubflow}
\end{figure}

We collected the publications of each matched faculty member in the periods of the faculty lists, 30{,}488 faculty--publication records and 26{,}053 unique publications; the publications of 2025--2026 were retrieved on September 26, 2026.
Because publishing norms differ across the areas of LIS, we kept journal articles, conference papers, books, and book chapters, including records without a Dimensions document type, and removed 7{,}931 records that are not full, peer-reviewed research reports, such as editorials, preprints, and short conference contributions (Figure~\ref{fig:pubflow}; Appendix~C gives the criteria).
An initial theme model also revealed two small themes outside LIS, astrophysics and quantum communication and computing, which came almost entirely from two faculty members; removing their 74 publications left an analysis corpus of 18{,}048 publications.
Of these, 2{,}114 have no abstract of at least ten words in Dimensions.
We left them out of the theme model, which places publications by their content (in a preliminary model, records represented by their title alone formed a theme of their own), but kept them in the publication counts and the venue analysis.

\subsection{Research Theme Modeling}
\label{sec:thememodel}
We identified research themes with BERTopic \parencite{grootendorst2022bertopic}, which clusters transformer-based document embeddings into interpretable groups.
Although BERTopic calls its clusters topics, we call them research themes, because their granularity is closer to that of domain-level areas of LIS.
We embedded the title and abstract of each publication, joined by the separation token, with SPECTER2 \parencite{singh2023scirepeval}, a model designed to represent scholarly documents, which yields a 768-dimensional vector per publication.
UMAP (Uniform Manifold Approximation and Projection) reduced the embeddings to ten dimensions; the themes change little with 5 to 50 dimensions (Appendix~D).
We then grouped the publications by agglomerative clustering with Ward linkage on these coordinates.
Ward linkage builds a single hierarchy of nested clusters, and cutting it at $K$ clusters gives $K$ themes.
Our analyses use the cut at $K=15$, and the cuts at $K=4$ and $K=30$ describe the field at a coarser and a finer resolution; because the levels nest, each of the four areas is a union of themes and each theme a union of finer themes (Table~D3 in Appendix~D).
We chose the clustering method and the three levels by comparing candidate cuts on term coherence, term diversity, agreement with the Fields of Research classification of Dimensions, and stability across subsamples and halves of the corpus (Appendix~D).
Each theme is represented by its most distinctive keywords, extracted with class-based TF-IDF (c-TF-IDF) and Maximal Marginal Relevance (MMR) to keep them diverse.
A large language model (Gemini 3.8 Flash) proposed a label, subtopics, and a brief summary for each theme from the titles of all its publications, and we revised the labels that did not fit the publications (see Uses and Roles of Generative AI).
For the research landscape map (Figure~\ref{fig:research_landscape}), UMAP reduced the embeddings to two dimensions; theme colors follow the position of each theme on the map, so neighboring themes have similar hues.

\subsection{Publication Venue Classification}
To describe the research of the schools also by where their faculty publish, we used the venue of each publication, including those without an abstract: the journal, book series, or conference series in which it appeared, with the editions of a conference merged into one venue.
We classified the venues into ten fields: Library and Information Science (LIS); Computer Science, including human--computer interaction; Education; Communication and Media; Health and Biomedicine; Management and Information Systems; Social Sciences; Humanities; Multidisciplinary Science; and Natural Sciences and Engineering.
A venue's field comes from its Web of Science subject categories or, for venues outside the Web of Science, from the Dimensions Fields of Research of its publications, each mapped by hand to one of the ten fields following the OECD Fields of Science and Technology \parencite{oecd2015frascati,kaliuzhna2024mapping}; Appendix~E gives the details.
Of the 18{,}048 publications, 16{,}688 have a venue field, and most of the other 1{,}360 (7.5\%) are books and chapters outside a series.

\subsection{Research Profiles}
\label{sec:profiles}
We built the research profiles of the schools from their faculty members, period by period.
For each period, a listed faculty member's profile is the share of their publications of that period in each of the 15 themes; faculty without a publication in the period have no profile for it.
A school's profile for a period is the mean of its faculty members' profiles, its overall profile is the mean of its period profiles, and a school type's profile is the mean of its schools' overall profiles.
Every faculty member thus counts equally within a school and period, every period within a school, and every school within a type, so that neither highly productive faculty, large schools, nor periods with more publications dominate a profile.
We compare each type's share of a theme with that of the average school, the mean profile of all 44 schools.
Venue profiles are built in the same way, with the share of a faculty member's publications in venues of each field in place of the themes.

\subsection{Statistical Analysis}
We tested whether the five school types differ in what their faculty members publish with Permutational Multivariate Analysis of Variance \parencite[PERMANOVA;][]{anderson2001new}, once for the 15 themes and once for the ten venue fields.
Each faculty member at a school is one unit, described by the shares of all their publications of 2013--2026 at that school in the themes or fields.
We measured the difference between two faculty members by the Hellinger distance, the Euclidean distance between the square roots of their shares \parencite{legendre2001ecologically}, which needs no adjustment for the many themes in which a faculty member has no publication and gives every profile the same weight.
Because school type is a property of the school, we used a nested design \parencite{anderson2003permutation}: PERMANOVA splits the variation among faculty members into parts between the types, between the schools within a type, and among the faculty members of a school, and it tests the types against the variation between schools within a type, with the $p$ value from 9{,}999 permutations of the type labels among whole schools.
We report the pseudo-$F$ statistic, the share of the variation between the types ($R^2$), and $p$, and we compared the ten pairs of types only after a significant test of all five (Fisher's protected least significant difference).
As a check, we repeated the tests with each school as one unit (Appendix~F).
To show how similar the schools are, we placed them on the first two principal components of their theme compositions and, separately, of their venue-field compositions, each school at the mean of its faculty members' square-root profiles, the space in which PERMANOVA compares the faculty members.

We measured the research diversity of a school as the effective number of themes \parencite{hill1973diversity}, the exponential of the Shannon entropy of its distribution over the themes, which ranges from 1 (all publications in one theme) to 15 (an even spread over all themes).
Because larger schools have more faculty members to spread over the themes, we compared the schools at an equal sample coverage of 92.5\% \parencite{chao2012coverage}, estimated with iNEXT \parencite{hsieh2016inext} with a faculty member in one period as the sampling unit, and compared the types with the Kruskal--Wallis test.
To follow the types over time, we computed each school's shares of the four research areas and of the venue fields, and its diversity, in each period.
For diversity over time we used the 30 finer themes, because a school has few publications in one period (a median of 41) and the 15 themes would join lines of work that the finer themes keep apart.
A type's value in a period is the mean over its schools, with a 95\% interval from resampling each type's schools with replacement (4{,}000 resamples), and its trend is the least-squares slope of these means over the midpoints of the periods, per decade, with an interval from the same resamples.
We also computed the shares and the diversity from each school's publications of the period, each counted once, so that faculty members with more publications weigh more, and Appendix~F compares diversity at equal numbers of faculty members.


\section{Results}

\subsection{Faculty and Publication Data}
Of the 44 schools, 33 are at R1 universities in the 2021 Carnegie Classification \citep{carnegie2021}, including all seven \textsc{Communication} schools (Table~\ref{tab:stats}).
The five faculty lists name 2{,}292 faculty members, 69.3\% of them tenured or tenure-track, and 1{,}602 (69.9\%) were matched to a Dimensions profile: 84.4\% of the tenured and tenure-track and 39.1\% of the non-tenure-track faculty, many of whom hold teaching-focused positions.
\textsc{Information} (71.4 faculty members per school) and \textsc{Computer} schools (60.5) are more than twice the size of \textsc{Education} schools (28.2).
Conference papers make up 35.5\% of the publications of \textsc{Information} and 29.7\% of \textsc{Computer} schools, against about a fifth in the other types (Table~\ref{tab:pubs}).
Because the types differ this much, every school counts equally in the profiles below.

\begin{table}[!htbp]
    \caption{Schools and Faculty by School Type}
    \centering
    \footnotesize
    \begin{tabular}{lrrrrrrrrrrr}
    \toprule
    & \multicolumn{5}{c}{Schools} & \multicolumn{3}{c}{Faculty} & \multicolumn{3}{c}{With Dimension IDs} \\
    \cmidrule(lr){2-6}\cmidrule(lr){7-9}\cmidrule(lr){10-12}
    Type & All & R1 & R2 & R3 & MS & All & TT & NTT & All & TT & NTT \\
    \midrule
    \textsc{Art\&Science}  & 6  & 3  & 1 & 0 & 2 & 223      & 146      & 77  & 127      & 105      & 22  \\
    \textsc{Communication} & 7  & 7  & 0 & 0 & 0 & 331      & 221      & 84  & 218      & 184      & 31  \\
    \textsc{Computer}      & 8  & 6  & 1 & 0 & 1 & 484      & 348      & 128 & 362      & 302      & 59  \\
    \textsc{Education}     & 9  & 6  & 2 & 1 & 0 & 254      & 168      & 61  & 167      & 137      & 21  \\
    \textsc{Information}   & 14 & 11 & 0 & 1 & 2 & 1{,}000  & 706      & 277 & 728      & 613      & 112 \\
    \midrule
    Total                  & 44 & 33 & 4 & 2 & 5 & 2{,}292  & 1{,}589  & 627 & 1{,}602  & 1{,}341  & 245 \\
    \bottomrule
    \end{tabular}

    \vspace{4pt}
    \begin{minipage}{0.95\linewidth}
    \footnotesize\textit{Note.} Schools by the 2021 Carnegie Classification of their university: R1 = very high and R2 = high research activity; R3 = doctoral/professional; MS = master's. Faculty = members of any of the five faculty lists (2014--2026), counted at each school that listed them. TT = tenured or tenure-track; NTT = non-tenure-track, by the last listed position; the 76 faculty members of unknown track (16 matched) are counted under All only. With Dimension IDs = matched to a Dimensions researcher profile.
    \end{minipage}
    \label{tab:stats}
\end{table}

\begin{table}[!htbp]
    \caption{Publications by School Type and Publication Type}
    \centering
    \footnotesize
    \begin{tabular}{lrrrrrrrrr}
    \toprule
    & & \multicolumn{2}{c}{Journal articles} & \multicolumn{2}{c}{Conference papers} & \multicolumn{2}{c}{Book chapters} & \multicolumn{2}{c}{Books} \\
    \cmidrule(lr){3-4}\cmidrule(lr){5-6}\cmidrule(lr){7-8}\cmidrule(lr){9-10}
    Type & Publications & $n$ & \% & $n$ & \% & $n$ & \% & $n$ & \% \\
    \midrule
    \textsc{Art\&Science}  & 809      & 540      & 66.7 & 152      & 18.8 & 111      & 13.7 & 6   & 0.7 \\
    \textsc{Communication} & 2{,}467  & 1{,}772  & 71.8 & 500      & 20.3 & 177      & 7.2  & 18  & 0.7 \\
    \textsc{Computer}      & 5{,}391  & 3{,}302  & 61.3 & 1{,}600  & 29.7 & 453      & 8.4  & 36  & 0.7 \\
    \textsc{Education}     & 1{,}395  & 937      & 67.2 & 273      & 19.6 & 171      & 12.3 & 14  & 1.0 \\
    \textsc{Information}   & 8{,}661  & 4{,}922  & 56.8 & 3{,}076  & 35.5 & 605      & 7.0  & 58  & 0.7 \\
    \midrule
    Total                  & 18{,}048 & 11{,}043 & 61.2 & 5{,}398  & 29.9 & 1{,}477  & 8.2  & 130 & 0.7 \\
    \bottomrule
    \end{tabular}

    \vspace{4pt}
    \begin{minipage}{0.95\linewidth}
    \footnotesize\textit{Note.} Analysis corpus, 2013--2026; \% = share of the type's publications. A publication with authors at schools of two types is counted under both, so the rows sum to more than the 18{,}048 unique publications. Conference papers include proceedings that Dimensions files as journals or book series, such as the ASIS\&T Annual Meeting and Lecture Notes in Computer Science (Appendix~C).
    \end{minipage}
    \label{tab:pubs}
\end{table}

\subsection{Research Themes}
\label{sec:researchthemes}
\begin{figure}[!htbp]
    \centering
    \includegraphics[width=0.9\linewidth]{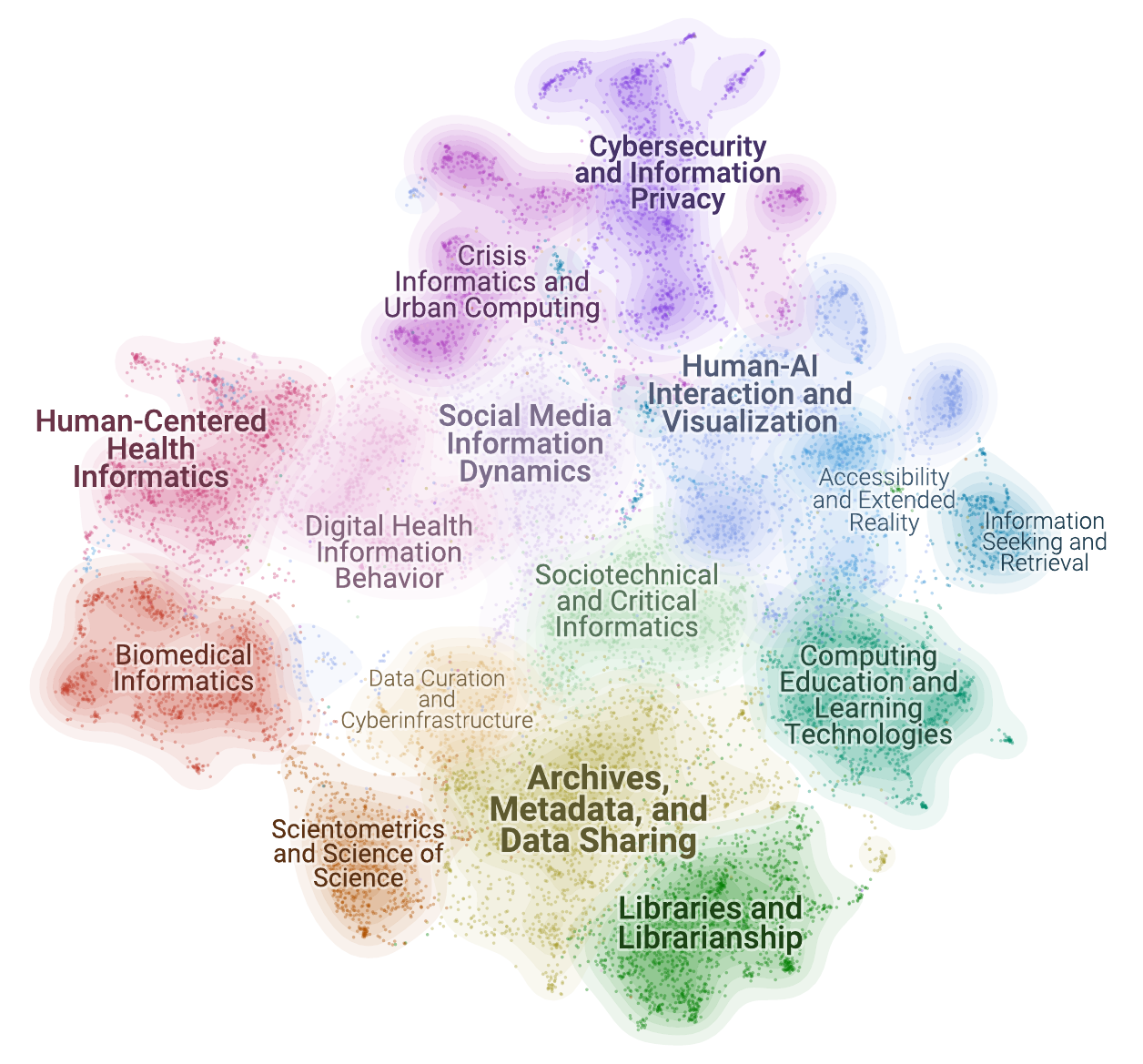}
    \caption{Research landscape of Library and Information Science in the United States from 2013 to 2026. Each point is a publication positioned by semantic similarity. Colors denote the 15 research themes, with similar hues for neighboring themes; dense regions and larger labels mark higher volume. }
    \label{fig:research_landscape}
\end{figure}

The 15{,}934 publications with an abstract fall into the 15 research themes, which form four areas at the coarsest level of the hierarchy (Table~\ref{tab:topic_labels_dataset} and Figure~\ref{fig:research_landscape}).

Human-Centered and Social Computing (36.4\% of the publications) studies how people work, learn, and communicate with computing and with one another online, from human-AI interaction, visualization, and social media to computing education, critical informatics, accessibility, and interactive search.
Libraries, Archives, and Scholarly Communication (29.2\%) holds the field's traditional subjects: archives and cultural heritage, metadata and research data, libraries and librarianship, citation analysis, and the curation of scientific data.
Health Informatics (21.1\%) spans clinical systems and care, health information behavior, and biomedical data.
Security, Networks, and Urban Computing (13.4\%) joins cybersecurity, networks, and privacy with disaster informatics and urban computing.
We use the four areas to group the themes in the research profiles and to follow the school types over time.

\begin{table}[!htbp]
    \caption{The Research Areas and Themes}
    \centering
    \singlespacing
    \footnotesize
    \setlength{\tabcolsep}{4pt}
    \begin{tabular}{>{\raggedright\arraybackslash\leftskip=1em}p{6.2cm}r>{\raggedright\arraybackslash}p{8.4cm}}
    \toprule
    \multicolumn{1}{l}{\textbf{Theme}} & \textbf{$n$} & \textbf{Representation} \\
    \midrule
    \multicolumn{3}{l}{\textit{Libraries, Archives, and Scholarly Communication} (4{,}647)} \\
    Archives, Metadata, and Data Sharing & 1{,}893 & archival, data, archives, metadata, collections, knowledge, curation, cultural, reuse \\
    \addlinespace[2pt]
    Libraries and Librarianship & 1{,}514 & librarians, school, literacy, librarianship, justice, books, librarian, patrons, communities \\
    \addlinespace[2pt]
    Scientometrics and Science of Science & 774 & citation, citations, articles, scholarly, disciplines, retracted, authorship, citing, career \\
    \addlinespace[2pt]
    Data Curation and Cyberinfrastructure & 466 & provenance, science, data, quantum, citizen, metadata, wildlife, workflows, quality \\
    \addlinespace
    \multicolumn{3}{l}{\textit{Human-Centered and Social Computing} (5{,}793)} \\
    Human-AI Interaction and Visualization & 1{,}330 & visual, visualizations, robots, intelligence, artificial, language, llms, tasks, humans \\
    \addlinespace[2pt]
    Social Media Information Dynamics & 1{,}283 & media, political, misinformation, twitter, fake, wikipedia, moderation, trolling, disinformation \\
    \addlinespace[2pt]
    Computing Education and Learning Technologies & 1{,}032 & students, children, programming, game, teachers, education, learners, classroom, courses \\
    \addlinespace[2pt]
    Sociotechnical and Critical Informatics & 1{,}030 & hci, workers, technologies, social, migrants, use, city, communities, icts \\
    \addlinespace[2pt]
    Accessibility and Extended Reality & 583 & vr, accessibility, blind, autistic, interaction, adults, touch, gestures, immersive \\
    \addlinespace[2pt]
    Information Seeking and Retrieval & 535 & search, retrieval, query, tasks, searchers, recommender, topic, sentiment, web \\
    \addlinespace
    \multicolumn{3}{l}{\textit{Health Informatics} (3{,}362)} \\
    Human-Centered Health Informatics & 1{,}322 & care, older, caregivers, healthcare, ehr, medication, dementia, adherence, support \\
    \addlinespace[2pt]
    Digital Health Information Behavior & 1{,}032 & health, covid, online, social, sexual, women, hiv, alcohol, stigma \\
    \addlinespace[2pt]
    Biomedical Informatics & 1{,}008 & disease, gene, cancer, rna, proteins, ontology, cells, data, prediction \\
    \addlinespace
    \multicolumn{3}{l}{\textit{Security, Networks, and Urban Computing} (2{,}132)} \\
    Cybersecurity and Information Privacy & 1{,}121 & privacy, security, cloud, cybersecurity, blockchain, iot, cyber, wireless, energy \\
    \addlinespace[2pt]
    Crisis Informatics and Urban Computing & 1{,}011 & network, drivers, disaster, spatial, model, urban, detection, vehicles, prediction \\
    \bottomrule
    \end{tabular}

    \vspace{4pt}
    \begin{minipage}{0.95\linewidth}
    \footnotesize\textit{Note.} Italic rows are the four areas of the coarsest level ($K=4$), with their number of publications. Representation gives the keywords that best represent each theme. The areas and themes are cuts of the same Ward linkage, so each area is a union of themes; Table~D3 in Appendix~D divides the themes into the 30 finer themes of the $K=30$ cut.
    \end{minipage}
    \label{tab:topic_labels_dataset}
\end{table}

\subsection{Research Profiles by School Type}
\label{sec:typeprofiles}
\begin{figure}[!htbp]
    \centering
    \includegraphics{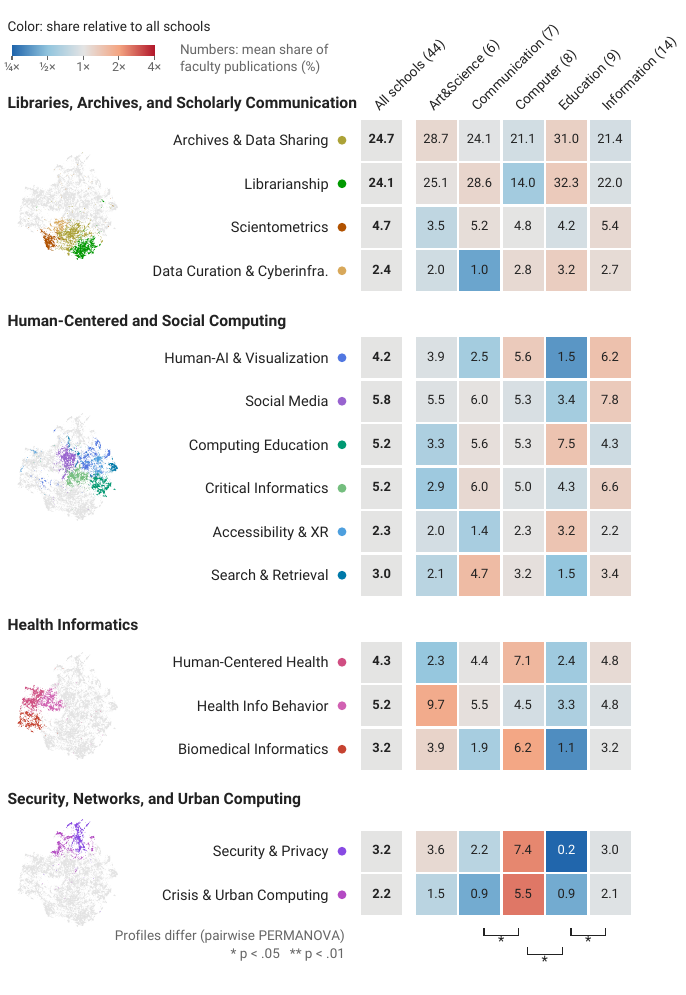}
    \caption{Research profiles of the five school types. Rows are grouped into the four areas; each small map highlights the area's themes on the research landscape of Figure~\ref{fig:research_landscape}. Each number is the type's mean share of faculty publications in the theme (\%), and the first column gives the share in the average school. Colors compare each type's share with that of the average school, from a quarter (blue) to four times (red). Brackets join the pairs of types whose faculty members' profiles differ in the pairwise PERMANOVAs.}
    \label{fig:school_type_profiles}
\end{figure}

Figure~\ref{fig:school_type_profiles} shows the research profile of each school type, based on 3{,}003 profiles of 1{,}321 faculty members over the five periods.

Libraries, Archives, and Scholarly Communication is the largest area in every type, from 42.6\% of faculty publications in \textsc{Computer} schools to 70.6\% in \textsc{Education} schools (56.1\% in the average school); the types differ in where the rest of their research goes.

\textsc{Computer} schools lean toward computing and technical research.
Security, Networks, and Urban Computing takes 12.9\% of their faculty publications, 2.4 times its share in the average school, and they lead all types in Human-Centered Health Informatics (7.1\%) and Biomedical Informatics (6.2\%), whereas Libraries and Librarianship falls to 14.0\%, little more than half its average share (24.1\%).
\textsc{Education} schools are the opposite case: their largest theme outside Libraries, Archives, and Scholarly Communication is Computing Education and Learning Technologies (7.5\%, the highest share of any type), in line with the education units they share a parent with, while Health Informatics (6.8\%) and Security, Networks, and Urban Computing (1.1\%) reach only half and a fifth of their average shares.
\textsc{Communication} schools publish more on Information Seeking and Retrieval than any other type (4.7\%, 1.6 times the average share), and \textsc{Art\&Science} schools nearly double the average share of Digital Health Information Behavior (9.7\% against 5.2\%), but both stay close to the average school on most themes.
\textsc{Information} schools stay between 0.83 and 1.47 times the average share on every theme, yet they lean toward Human-Centered and Social Computing more than any other type (30.5\%, against 25.8\% in the average school).

The theme profiles of the 1{,}321 faculty members differ by school type (PERMANOVA pseudo-$F = 2.58$, $R^2 = .022$, $p = .041$), but type accounts for only 2.2\% of the variation among faculty members and the schools within a type for 8.4\%; the remaining 89.4\% lies among the faculty members of the same school.
In the pairwise comparisons, \textsc{Computer} schools differ from \textsc{Communication} ($p = .020$) and \textsc{Education} schools ($p = .013$), and \textsc{Education} schools from \textsc{Information} schools ($p = .042$).

\subsection{Research Diversity by School Type}
\label{sec:diversity}
\begin{figure}[!htbp]
    \centering
    \includegraphics{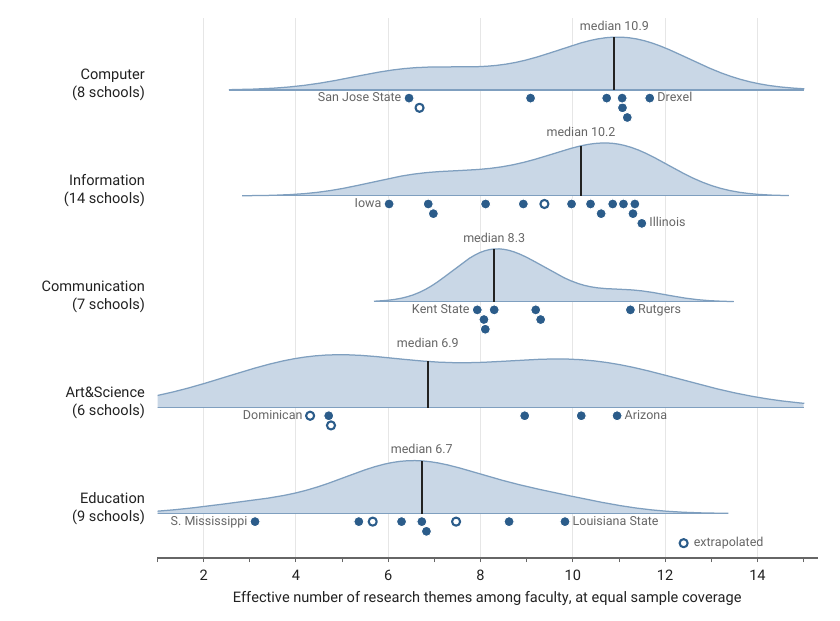}
    \caption{Research diversity of the schools by school type. Each dot is a school. Open dots are schools whose estimate is extrapolated beyond their sample. The shaded area is a kernel density estimate.}
    \label{fig:school_diversity}
\end{figure}

Figure~\ref{fig:school_diversity} shows the research diversity of the schools at equal sample coverage.
\textsc{Computer} (median 10.9 themes) and \textsc{Information} schools (10.2) are the most diverse, followed by \textsc{Communication} (8.3), \textsc{Art\&Science} (6.9), and \textsc{Education} schools (6.7), and the types differ (Kruskal--Wallis $H = 12.2$, $df = 4$, $p = .016$).
Within every type, the most and the least diverse school differ by 3.3 to 6.7 themes, and the six \textsc{Art\&Science} schools split into two groups of three, one between 4.3 and 4.8 themes and one between 9.0 and 11.0.
Without the coverage adjustment, the medians keep the same order (Appendix~F).

\subsection{Publication Venues by School Type}
\label{sec:venueprofiles}


\begin{figure}[!htbp]
    \centering
    \includegraphics{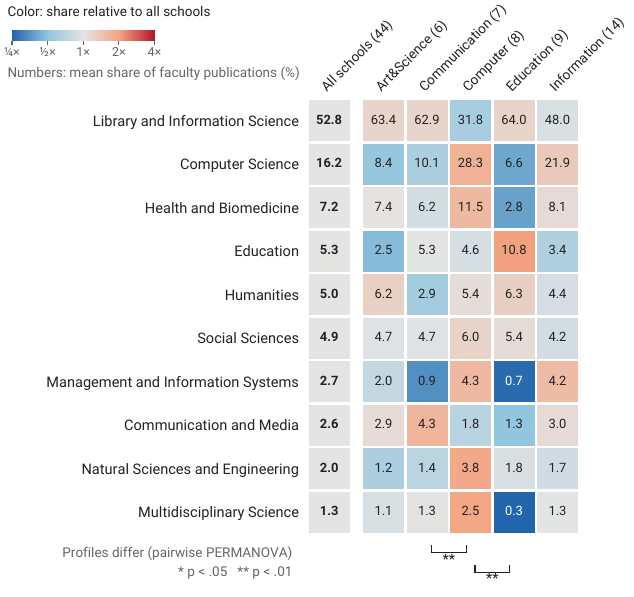}
    \caption{Publication venue profiles of the five school types. Each number is the type's mean share of faculty publications in venues of the field (\%); the first column, colors, and brackets are as in Figure~\ref{fig:school_type_profiles}.}
    \label{fig:school_type_venue_profiles}
\end{figure}

\begin{table}[!htbp]
    \caption{The 10 Venues With the Most Publications in Each School Type}
    \centering
    \singlespacing
    \scriptsize
    \setlength{\tabcolsep}{4pt}
    \begin{tabular}{r>{\raggedright\arraybackslash}p{9.2cm}lrrr}
    \toprule
    & & & & \multicolumn{2}{c}{Share (\%)} \\
    \cmidrule(lr){5-6}
    & Venue & Field & $N$ & Publications & Faculty mean \\
    \midrule
    \multicolumn{6}{l}{\textsc{Art\&Science} (6 schools; 731 publications)} \\
    1 & Journal of the Association for Information Science and Technology & LIS & 30 & 4.1 & 3.4 \\
    2 & ASIS\&T Annual Meeting & LIS & 18 & 2.5 & 2.5 \\
    3 & Journal of Education for Library and Information Science & LIS & 16 & 2.2 & 5.1 \\
    4 & The Library Quarterly & LIS & 14 & 1.9 & 1.6 \\
    5 & Cataloging \& Classification Quarterly & LIS & 13 & 1.8 & 4.0 \\
    6 & Library \& Information Science Research & LIS & 13 & 1.8 & 2.7 \\
    7 & Journal of Librarianship and Information Science & LIS & 11 & 1.5 & 3.7 \\
    8 & ALISE Annual Conference & LIS & 10 & 1.4 & 1.6 \\
    9 & Archival Science & LIS & 9 & 1.2 & 1.5 \\
    10 & Library Trends & LIS & 9 & 1.2 & 1.2 \\
    \addlinespace[2pt]
    \multicolumn{6}{l}{\textsc{Communication} (7 schools; 2{,}322 publications)} \\
    1 & Journal of the Association for Information Science and Technology & LIS & 90 & 3.9 & 4.6 \\
    2 & ASIS\&T Annual Meeting & LIS & 90 & 3.9 & 4.6 \\
    3 & ALISE Annual Conference & LIS & 66 & 2.8 & 3.8 \\
    4 & Journal of Documentation & LIS & 56 & 2.4 & 3.3 \\
    5 & Library \& Information Science Research & LIS & 48 & 2.1 & 2.7 \\
    6 & Journal of Education for Library and Information Science & LIS & 43 & 1.9 & 3.0 \\
    7 & The Library Quarterly & LIS & 41 & 1.8 & 2.1 \\
    8 & Journal of Librarianship and Information Science & LIS & 36 & 1.6 & 1.8 \\
    9 & iConference & LIS & 36 & 1.6 & 1.5 \\
    10 & Information Processing \& Management & LIS & 33 & 1.4 & 1.9 \\
    \addlinespace[2pt]
    \multicolumn{6}{l}{\textsc{Computer} (8 schools; 4{,}947 publications)} \\
    1 & CHI Conference on Human Factors in Computing Systems & CS & 167 & 3.4 & 2.0 \\
    2 & Journal of the Association for Information Science and Technology & LIS & 118 & 2.4 & 2.8 \\
    3 & Proceedings of the ACM on Human-Computer Interaction & CS & 112 & 2.3 & 1.6 \\
    4 & PLOS ONE & MULTI & 98 & 2.0 & 0.9 \\
    5 & ASIS\&T Annual Meeting & LIS & 58 & 1.2 & 1.8 \\
    6 & Scientometrics & LIS & 50 & 1.0 & 0.4 \\
    7 & Hawaii International Conference on System Sciences & MGMT & 44 & 0.9 & 2.3 \\
    8 & IEEE International Conference on Big Data & CS & 44 & 0.9 & 0.9 \\
    9 & International AAAI Conference on Web and Social Media & CS & 44 & 0.9 & 0.5 \\
    10 & Journal of Informetrics & LIS & 43 & 0.9 & 0.3 \\
    \addlinespace[2pt]
    \multicolumn{6}{l}{\textsc{Education} (9 schools; 1{,}255 publications)} \\
    1 & ALISE Annual Conference & LIS & 63 & 5.0 & 3.2 \\
    2 & ASIS\&T Annual Meeting & LIS & 33 & 2.6 & 3.6 \\
    3 & Journal of Documentation & LIS & 33 & 2.6 & 2.8 \\
    4 & Journal of Education for Library and Information Science & LIS & 30 & 2.4 & 3.7 \\
    5 & Science and Children & EDU & 30 & 2.4 & 1.4 \\
    6 & Journal of the Association for Information Science and Technology & LIS & 29 & 2.3 & 2.2 \\
    7 & Annual Conference of CAIS & LIS & 29 & 2.3 & 1.8 \\
    8 & Public Library Quarterly & LIS & 27 & 2.2 & 3.4 \\
    9 & Library \& Information Science Research & LIS & 26 & 2.1 & 3.5 \\
    10 & SLIS Connecting & LIS & 23 & 1.8 & 2.5 \\
    \addlinespace[2pt]
    \multicolumn{6}{l}{\textsc{Information} (14 schools; 8{,}079 publications)} \\
    1 & CHI Conference on Human Factors in Computing Systems & CS & 538 & 6.7 & 2.5 \\
    2 & Proceedings of the ACM on Human-Computer Interaction & CS & 323 & 4.0 & 2.1 \\
    3 & Journal of the Association for Information Science and Technology & LIS & 202 & 2.5 & 4.3 \\
    4 & ASIS\&T Annual Meeting & LIS & 200 & 2.5 & 2.7 \\
    5 & iConference & LIS & 117 & 1.4 & 1.4 \\
    6 & Hawaii International Conference on System Sciences & MGMT & 107 & 1.3 & 1.0 \\
    7 & International AAAI Conference on Web and Social Media & CS & 94 & 1.2 & 0.4 \\
    8 & The Library Quarterly & LIS & 73 & 0.9 & 1.3 \\
    9 & ALISE Annual Conference & LIS & 71 & 0.9 & 2.1 \\
    10 & ACM Conference on Computer Supported Cooperative Work and Social Computing & CS & 66 & 0.8 & 0.5 \\
    \bottomrule
    \end{tabular}

    \vspace{4pt}
    \begin{minipage}{0.95\linewidth}
    \footnotesize\textit{Note.} $N$ is the number of the type's publications in the venue.
    Publications is the venue's share of all the type's publications with a venue field.
    Faculty mean is the type's mean share of faculty publications in the venue.
    LIS = Library and Information Science; CS = Computer Science; EDU = Education; MGMT = Management and Information Systems; MULTI = Multidisciplinary Science.
    \end{minipage}
    \label{tab:school_type_top_venues}
\end{table}

Figure~\ref{fig:school_type_venue_profiles} and Table~\ref{tab:school_type_top_venues} show the venue profiles of the school types, based on 16{,}688 publications with a venue field.
In the average school, LIS venues hold 52.8\% of faculty publications and Computer Science venues 16.2\%, and LIS venues hold the largest share in every type, from 31.8\% in \textsc{Computer} schools to 64.0\% in \textsc{Education} schools.

\textsc{Computer} schools have the highest shares of any type in Computer Science (28.3\%, 1.7 times the average share), Health and Biomedicine (11.5\%, 1.6 times), Natural Sciences and Engineering (3.8\%), and Multidisciplinary Science venues (2.5\%).
\textsc{Education} schools have the highest share in Education venues (10.8\%, 2.0 times) and the lowest in Health and Biomedicine (2.8\%), Computer Science (6.6\%), Management and Information Systems (0.7\%), and Multidisciplinary Science (0.3\%).
\textsc{Communication} schools have the highest share in Communication and Media venues (4.3\%, 1.7 times) and publish less than the average school in Computer Science venues (10.1\%).
\textsc{Art\&Science} schools publish 63.4\% of their work in LIS venues and about half the average share in Computer Science (8.4\%) and Education venues (2.5\%).
\textsc{Information} schools stay closest to the average school, between 0.6 and 1.6 times the average share in every field, with more than the average in Computer Science (21.9\%) and Management and Information Systems venues (4.2\%).
The Journal of the Association for Information Science and Technology and the ASIS\&T Annual Meeting are among the ten leading venues of every type (Table~\ref{tab:school_type_top_venues}).
The CHI Conference on Human Factors in Computing Systems leads the lists of \textsc{Computer} and \textsc{Information} schools, whereas nearly all ten venues of \textsc{Art\&Science}, \textsc{Communication}, and \textsc{Education} schools are LIS journals and conferences.

The venue profiles of the 1{,}329 faculty members also differ by school type (PERMANOVA pseudo-$F = 4.10$, $R^2 = .053$, $p = .010$), with 12.6\% of the variation between the schools within a type and 82.0\% among the faculty members of the same school.
In the pairwise comparisons, \textsc{Computer} schools differ from \textsc{Communication} ($p = .004$) and \textsc{Education} schools ($p = .003$); the difference between \textsc{Education} and \textsc{Information} schools falls short of significance ($p = .054$).
Type thus accounts for more of the variation in the venue fields (5.3\%) than in the research themes (2.2\%), and the tests with each school as one unit agree for both (Appendix~F).

\subsection{Similarity of Schools Within and Across Types}
\label{sec:similarity}
\begin{figure}[!htbp]
    \centering
    \includegraphics{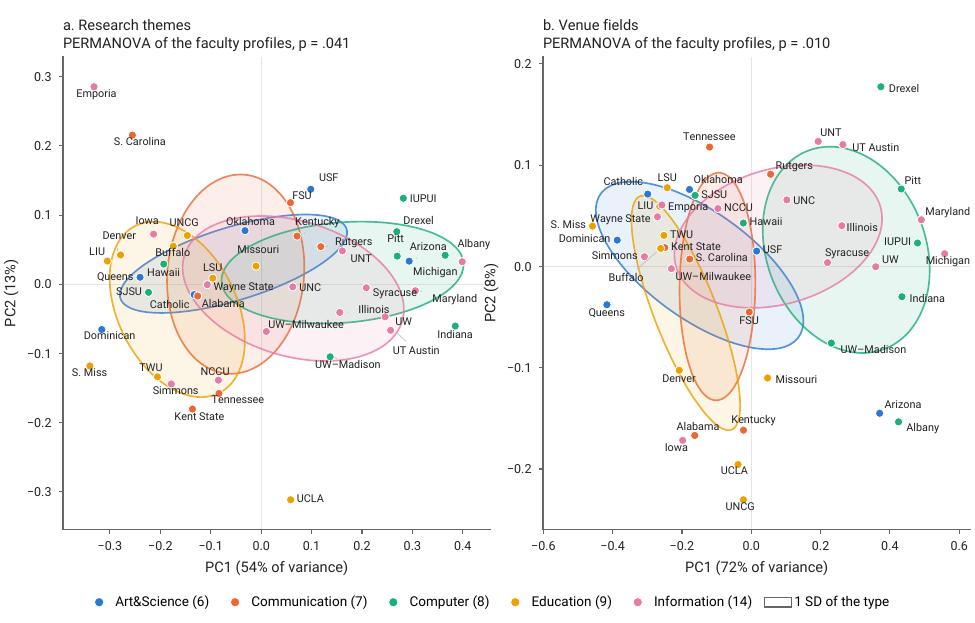}
    \caption{The 44 schools on the first two principal components of (a) their compositions over the 15 research themes and (b) their compositions over the ten venue fields. Each dot is a school, colored by its type; ellipses cover one standard deviation of each type's schools around their centroid.}
    \label{fig:school_similarity_pca}
\end{figure}

Figure~\ref{fig:school_similarity_pca} places the 44 schools by the similarity of their theme compositions (a) and their venue-field compositions (b); each ellipse shows how widely the schools of a type spread around their center.

A school's themes and its venues put it on the same main axis.
In both panels, the first component runs from libraries and archives to computational and technical research, and it accounts for 54\% of the variance among schools in the themes and 72\% in the venues.
Libraries and Librarianship and Archives, Metadata, and Data Sharing load at the library end of (a), and Human-AI Interaction and Visualization and Cybersecurity and Information Privacy lead the themes at the computational end; in (b), LIS venues load at the library end and Computer Science venues at the other.
On the first component, \textsc{Computer} schools lie furthest toward the computational end in both panels, followed by \textsc{Information} schools, and \textsc{Education} schools lie furthest toward the library end, with \textsc{Art\&Science} and \textsc{Communication} schools in between.
The ellipses overlap widely, and some schools lie far from the rest of their type.

\subsection{Changes Over Time}
\label{sec:trends}
\begin{figure}[!htbp]
    \centering
    \includegraphics{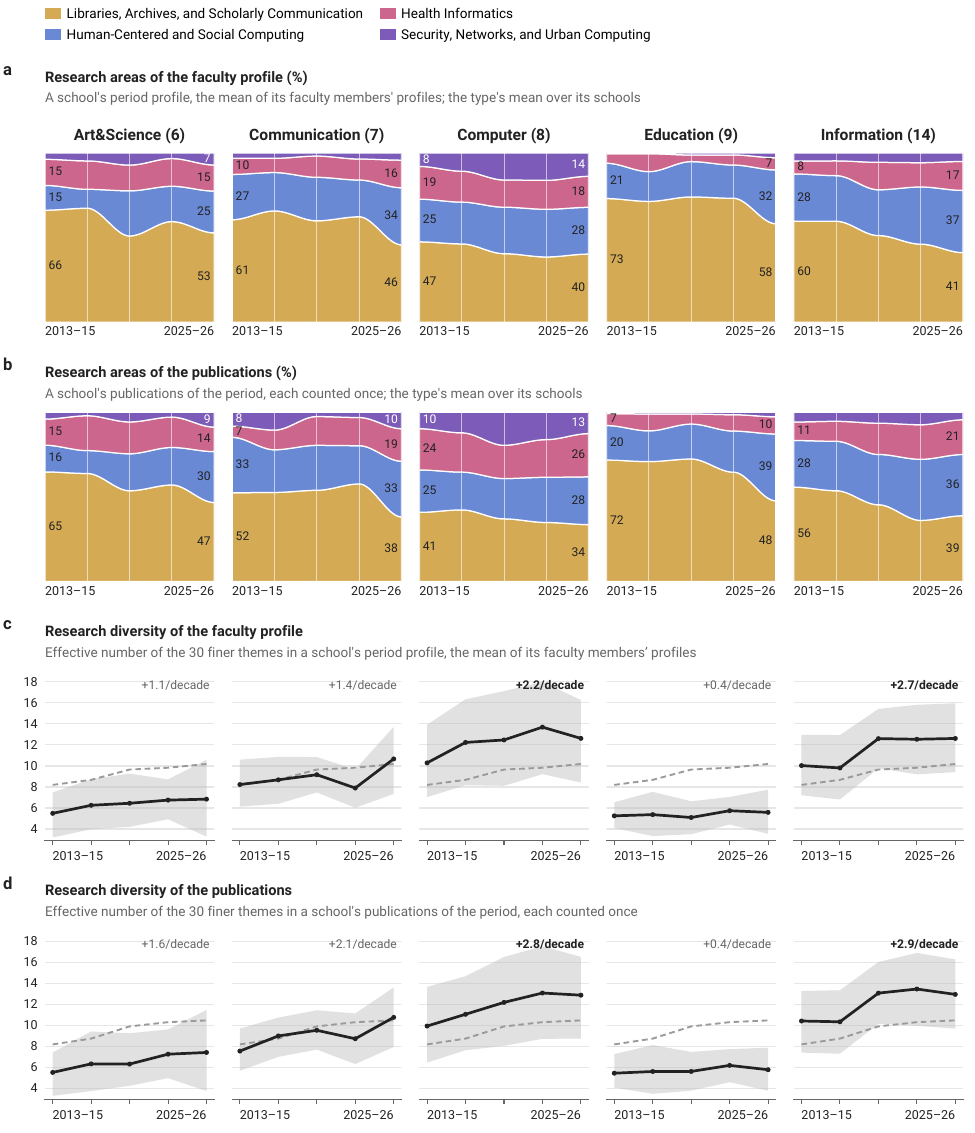}
    \caption{Research areas (a, b) and research diversity (c, d) of each school type in the five periods, from the faculty profiles (a, c) and from the publications, each counted once (b, d). In (a) and (b), each river shows the type's mean share (\%) of the four areas, which sum to 100\% in every period; each area has a softer tone of the color of its largest theme in Figure~\ref{fig:research_landscape}. In (c) and (d), research diversity on the 30 finer themes is shown as the type's mean (black line), its 95\% interval from resampling the schools (band), and the mean of all 44 schools (dashed).}
    \label{fig:area_trends}
\end{figure}

\begin{figure}[!htbp]
    \centering
    \includegraphics{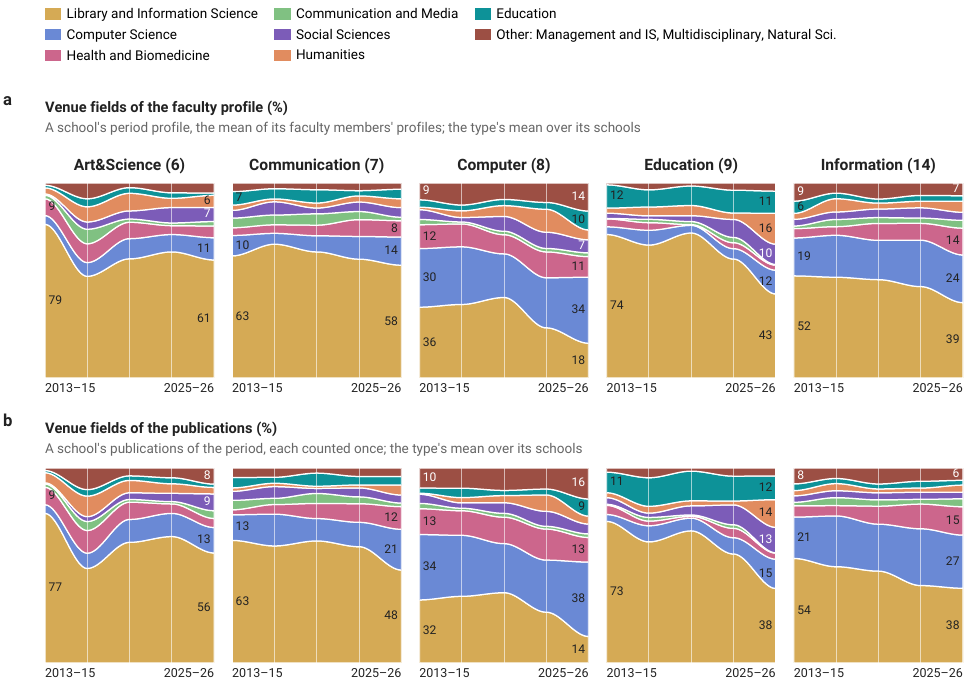}
    \caption{Publication venues of the school types over the five periods, from the faculty profiles (a) and from the publications (b), as in Figure~\ref{fig:area_trends}. Other combines Management and Information Systems, Multidisciplinary Science, and Natural Sciences and Engineering; Library and Information Science, Computer Science, and Health and Biomedicine have the colors of the matching research areas. Publications without a venue field (7.7\%) are left out.}
    \label{fig:venue_trends}
\end{figure}

Figures~\ref{fig:area_trends} and~\ref{fig:venue_trends} show the research areas, research diversity, and venue fields of each school type in the five periods, computed from the faculty profiles and from the publications.
We report the faculty profiles, with changes as slopes per decade and their 95\% intervals.

\paragraph{Research areas.} In the average school, the share of Libraries, Archives, and Scholarly Communication fell from 61.2\% of faculty publications in 2013--2015 to 46.5\% in 2025--2026, by 12.4 percentage points per decade ($[-18.0, -7.0]$), while Human-Centered and Social Computing rose from 24.0\% to 32.0\% ($+6.9$ $[2.4, 11.8]$), Health Informatics from 10.8\% to 15.1\% ($+3.4$ $[0.6, 6.2]$), and Security, Networks, and Urban Computing from 4.0\% to 6.4\% ($+2.1$ $[0.3, 4.5]$).
The decline of Libraries, Archives, and Scholarly Communication is largest in \textsc{Information} schools ($-17.4$) and also clear in \textsc{Art\&Science} schools ($-12.0$), and \textsc{Education} schools have the largest share of the area in every period.
\textsc{Computer} schools moved toward Security, Networks, and Urban Computing (from 7.9\% to 13.8\%; $+6.1$ $[0.2, 15.6]$), \textsc{Information} schools toward Human-Centered and Social Computing ($+8.3$ $[1.8, 14.9]$) and Health Informatics (from 7.7\% to 17.1\%; $+8.5$ $[3.6, 13.6]$), \textsc{Art\&Science} schools toward Human-Centered and Social Computing ($+10.3$ $[3.5, 17.0]$), and \textsc{Communication} schools toward Health Informatics ($+5.9$ $[2.6, 8.7]$).

\paragraph{Research diversity.} On the 30 finer themes, the diversity of the average school rose from 8.2 themes in 2013--2015 to 10.2 in 2025--2026 ($+1.8$ per decade $[0.9, 2.8]$; Figure~\ref{fig:area_trends}c).
The rise is clear in \textsc{Information} (from 10.0 to 12.6; $+2.8$ $[1.0, 4.8]$) and \textsc{Computer} schools (from 10.3 to 12.6; $+2.2$ $[0.3, 3.7]$), where the number of faculty profiles per school also grew (from 14.4 to 23.0 and from 15.5 to 21.0).
In \textsc{Communication}, \textsc{Art\&Science}, and \textsc{Education} schools, diversity rose more slowly, and the intervals include zero.
At equal numbers of faculty members, the diversity of the average school rose by 12\% rather than 24\%, and no type's interval excludes zero (Appendix~F).

\paragraph{Publication venues.} In the average school, the share of LIS venues fell from 59.0\% to 41.8\% ($-13.3$ percentage points per decade $[-19.1, -7.9]$), with intervals that exclude zero in every type but \textsc{Communication}, while the shares of Computer Science ($+4.0$ $[0.1, 7.7]$), Social Sciences ($+3.3$ $[1.1, 5.8]$), and Humanities venues ($+3.4$ $[0.5, 7.4]$) rose.
The fields that gained share differ by type and seldom match its partner discipline.
In the field closest to each type, the intervals include zero: Computer Science venues in \textsc{Computer} schools ($+1.0$ $[-14.4, 12.4]$), Education venues in \textsc{Education} schools ($+0.6$ $[-9.9, 13.2]$), and Communication and Media venues in \textsc{Communication} schools ($-2.4$ $[-6.5, 1.6]$).
Instead, Social Sciences venues rose in \textsc{Art\&Science} ($+6.0$ $[0.8, 12.2]$) and \textsc{Education} schools ($+7.8$ $[0.6, 16.1]$), Humanities venues in \textsc{Computer} schools ($+5.7$ $[1.8, 11.5]$), Computer Science venues in \textsc{Education} schools (from 4.0\% to 12.2\%), and Health and Biomedicine venues in \textsc{Communication} ($+4.5$ $[1.1, 8.1]$) and \textsc{Information} schools ($+7.5$ $[1.7, 15.4]$).

\subsection{Interactive Exploration}
\label{sec:liscape}
\begin{figure}[!htbp]
    \centering
    \includegraphics[width=\linewidth]{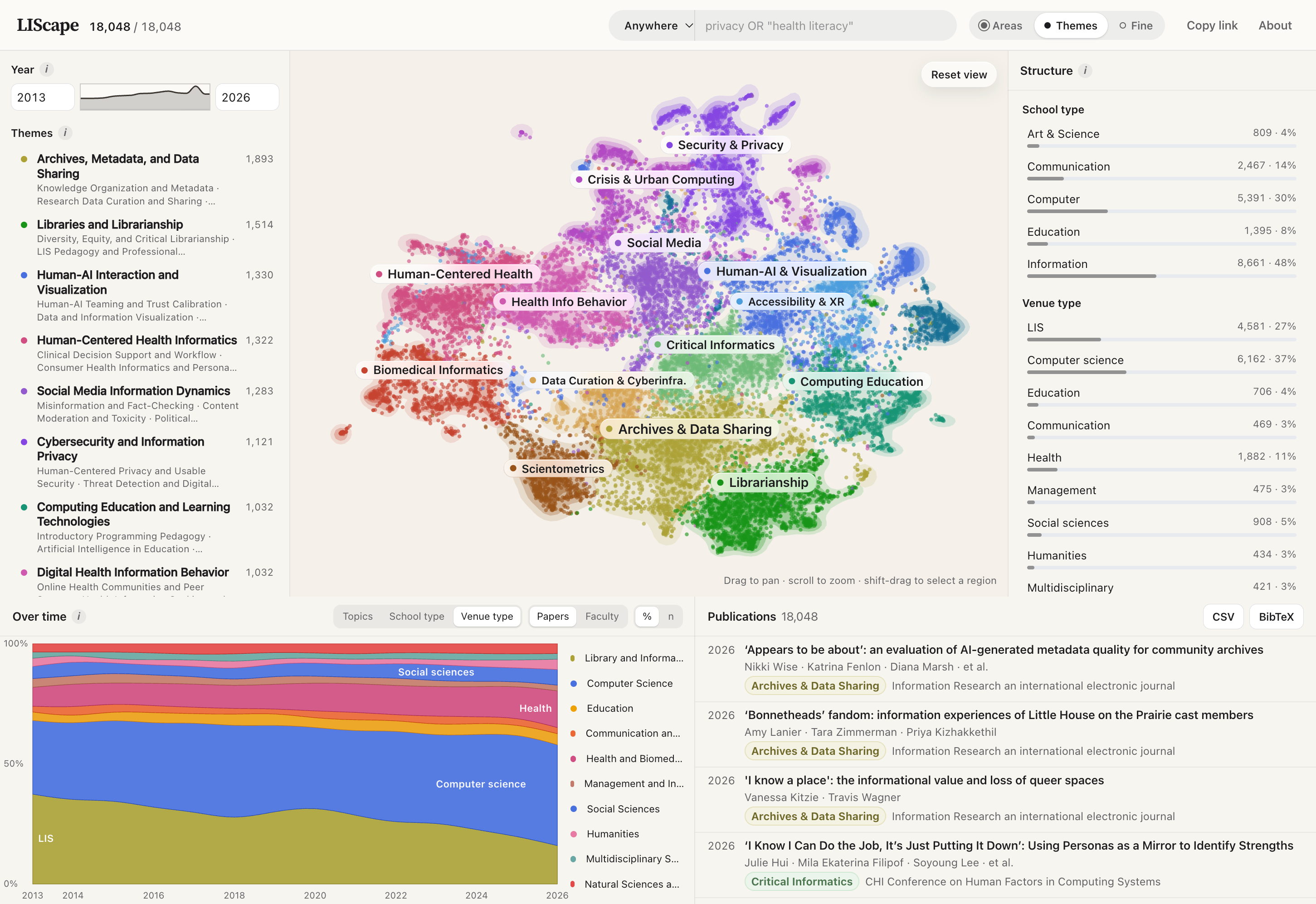}
    \caption{LIScape. Each point is a publication, placed by semantic similarity and colored by research theme; the year range, theme list, school-type and venue-field bars, timeline, and publication list all describe the same selection.}
    \label{fig:liscape}
\end{figure}

The comparisons above cannot cover every school, venue, or span of years that readers may ask about, so we built LIScape, an interactive view of the same corpus and theme model (Figure~\ref{fig:liscape}; \url{https://jiangenhe.com/liscape/}).

A reader selects publications by any combination of year, theme, school type, school, venue field, a rectangle on the map, and a search of titles, abstracts, authors, venues, and schools; the map, the bars, the timeline, and the list of publications then show that selection together.
The timeline can split the selection by theme, school type, or venue field and count it by publication or with equal weight for each faculty member, and the map can be read at the four areas, 15 themes, or 30 finer themes.

\section{Discussion}
The research portfolios of U.S.\ LIS schools differ by organizational structure, in the themes they emphasize and in how widely they spread across themes (RQ1), and the venues of their faculty differ along the same lines, more strongly (RQ2).
Both differences are small next to the variation among the faculty members of the same school.
From 2013 to 2026, the schools of every type moved away from the library core of LIS, along routes that differed by type but seldom led to the partner discipline (RQ3).

\paragraph{The Research Landscape of LIS Faculty}
Seen through the publications of its faculty, LIS has the core subjects of earlier structures of the field, with different weights.
Librarianship, citation analysis, and information seeking and retrieval, the core of the structures found in LIS publications \citep{Milojevi2011,astrom2007changes,zhao2014knowledge,han2020evolution,jarvelin2022lis} and dissertations \citep{Sugimoto2011}, are all themes here.
Three differences stand out.
First, information retrieval, the dominant domain in 14{,}035 LIS journal articles of 1996--2019 \citep{han2020evolution}, is the second-smallest theme here (3.4\% of the publications), whereas Libraries and Librarianship, which formed no top topic cluster there after 2000--2005, is the second-largest (9.5\%).
Second, Health Informatics (21.1\%) has no counterpart in the branches of \citet{Milojevi2011}, the dimensions of \citet{Wu2025}, or the seven communities in the research interests of iSchool faculty in 2013 \citep{holmberg2013conceptual}.
Third, Human-Centered and Social Computing, our largest area (36.4\%), has no category in the content analysis of LIS journal articles by \citet{jarvelin2022lis}, where the nearest subtopic, interactive retrieval, held 3.0\% of the articles in 2015.

The main reason is that most earlier structures were drawn from LIS journals, whose authors and LIS faculty overlap only in part.
First, many authors in LIS journals come from other disciplines: Computer Science made 21\% and Business and Economics 16\% of the contributions to 31 LIS journals in 2015, up from 9\% and 6\% in 1995 \citep{vakkari2023association}, and information retrieval, information seeking, and scientific communication in these journals depend heavily on such contributions \citep{Vakkari2022Disciplinary}.
Second, LIS faculty publish much of their work beyond LIS journals, especially in computing and health venues: in the average school 47.2\% of faculty publications appear outside LIS venues (see Publication Venues by School Type), and only 13.4\% of the publications in Human-Centered and Social Computing and 8.9\% of those in Health Informatics appear in LIS venues, against 70.1\% of those in Libraries, Archives, and Scholarly Communication.
Studies of faculty come closer to our structure: human--computer interaction was the most frequent research interest of iSchool faculty in 2013 \citep{holmberg2013conceptual} and one of the four most common research areas of U.S.\ LIS faculty, with a rising share \citep{Zhu2016}, and iSchool faculty placed rising emphasis on the intersection of information, technology, and people \citep{zuo2017state}.
Which structure a study finds thus depends on whose research it observes.

\paragraph{Organizational Structure, Research Portfolios, and Venues}
The differences between the school types are significant but modest: type accounts for 2.2\% of the variation among faculty members in the research themes and 5.3\% in the venue fields, far less than the variation among the faculty of each school.
The differences run mainly along one axis, from libraries and archives to computational and technical research, with \textsc{Computer} and \textsc{Education} schools at its two ends (Figure~\ref{fig:school_similarity_pca}).
The library core is shared: Libraries and Librarianship and Archives, Metadata, and Data Sharing are the two largest themes in every type, together from 35.1\% of faculty publications in \textsc{Computer} schools to 63.3\% in \textsc{Education} schools.

Where faculty members publish separates the types more than what they study.
Part of this follows from the themes, because the computing and health research of LIS faculty appears mostly outside LIS venues, but the stronger separation suggests that venues also reflect the scholarly communities that the faculty of a unit address and the standards by which their work is evaluated.
\citet{astrom2008formalizing} found a similar connection in Nordic LIS, where academic affiliation was linked to research orientation, to the media for communicating research, and to the extent to which research is evaluated by intra-disciplinary standards.

Our framework points to faculty composition as one link between organizational type and research.
iSchools draw their faculty from many intellectual domains \citep{Wiggins2012}, and of the U.S.\ LIS faculty hired from 1975 to 2015, 60\% held a doctorate in LIS, followed by computer science and education, with a rising share in computer science \citep{Zhu2016}.
Computer science and education are the partner disciplines of the two types at the ends of the axis, as would be expected if units recruit faculty from the discipline they are placed with, although we did not observe the backgrounds of the faculty.

\paragraph{Computing and the Fragmentation of LIS}
Earlier studies have tied the fragmentation of LIS to the contributions of other disciplines, computer science in particular, whose contributions to LIS journals are strongly differentiated by topic and method \citep{Vakkari2022Disciplinary,vakkari2023association}.
\citet{vakkari2024characterizes} traces this fragmentation to the 1960s and 1970s, when computer science and business established programs and research in information retrieval and bibliometrics, and, in the terms of \citet{whitley2000intellectual}, characterizes the LIS context, library and information services, and information seeking as fragmented adhocracies, and information retrieval and scientometrics as technologically integrated bureaucracies.

In the schools, computing appears elsewhere.
\textsc{Computer} schools publish the largest share of any type in Computer Science venues and the smallest in LIS venues, but their shares of Information Seeking and Retrieval (3.2\%) and of Scientometrics and Science of Science (4.8\%) are close to those of the average school (3.0\% and 4.7\%).
They differ instead in Security, Networks, and Urban Computing and in clinical and biomedical informatics, far from the retrieval and bibliometrics through which computer science first entered LIS.
Within the schools, the influence of computing thus appears less as computer scientists working on LIS problems than as the faculty of LIS units working on computing problems and publishing them outside LIS venues.
This division runs through the schools as much as between them: \textsc{Information}, \textsc{Art\&Science}, and \textsc{Computer} schools each span most of the axis from libraries to computing, while \textsc{Communication} and \textsc{Education} schools reach from its library end to about its middle (Figure~\ref{fig:school_similarity_pca}a).
Organizational type does not separate library schools from computing schools; it shifts the balance between library and computational research within schools that mostly do both.

\paragraph{Variation Within School Types}
One reading of the lean of \textsc{Computer} schools is resource dependence \citep{pfeffer1978external}: units placed with computer science gain access to its infrastructure, funding, and hiring networks, which favor computational research.
The variation within the types limits this reading.
Three of the eight \textsc{Computer} schools, the University of Hawaii--Manoa, San Jos\'e State University, and the University of Wisconsin--Madison, devote less of their faculty publications to Security, Networks, and Urban Computing and to clinical and biomedical informatics (0\%, 9.1\%, and 11.2\%) than most \textsc{Information} schools (median 14.2\%), whereas the other five devote between 23.1\% and 52.9\%.
All three changed their parent unit during the study window, each spending 4.3 to 6.7 of its 13.75 years under another type (Table~B1 in Appendix~B), and only Wisconsin--Madison moved clearly toward the computational end after the change, its share of Libraries, Archives, and Scholarly Communication falling from 77.2\% of faculty publications in 2016--2018 to 42.7\% in 2025--2026.
A school's research seems to follow the faculty it has more than the unit it reports to, and the faculty change slowly, through hiring, departures, and retirements.
If hiring follows curricular and programmatic needs, as noted in the Literature Review, a new parent unit would change a school's research only as fast as it changes the school's teaching programs and hiring.

\paragraph{Shared Change Across School Types}
The decline of the library core continues a long trend in LIS journals: library and information service activities fell from 27.2\% of the articles in 1985 to 13.9\% in 2015 \citep{jarvelin2022lis}, library-focused publications declined after 2011 while research centered on people and on algorithms grew \citep{Wu2025}, and social media and data science rose from about 5\% of the articles in 2012 to more than a third in 2018 \citep{ma2021evolution}.
The fall of LIS venues likewise fits the permeability of the field, since in 2010 nearly 60\% of the authors who published in LIS also published in another discipline \citep{Larivire2012}.
What these field-level studies cannot show is that the change differed by setting: the library core declined most in \textsc{Information} schools, \textsc{Education} schools kept the largest share of it in every period, and in the faculty profiles it remained the largest area in every type and period (Figure~\ref{fig:area_trends}a).
Because the shift seldom followed the partner discipline, it looks more like a development of the field as a whole than a pull from the parent unit.
The rise in diversity was clearest in \textsc{Computer} and \textsc{Information} schools, where the number of faculty profiles per school also grew, and it is smaller at equal numbers of faculty members (Appendix~F), so part of this broadening may come from adding faculty rather than from faculty changing their research.

\paragraph{Implications for Studies of the Field and for LIS Schools}
For studies of the LIS field, the results show that a landscape of the field depends on whose research it observes and averages over settings that differ, so studies could state which researchers and settings their corpus represents.
For LIS schools, the results suggest that a parent unit is not a quick route to a research portfolio: organizational type accounts for a small share of the variation among faculty members, and two of the three \textsc{Computer} schools nearest the library end kept their research through a change of parent unit.
Schools that move into units with data science or computer science, as the University of North Carolina at Chapel Hill and the University of Texas at Austin did in 2026, may therefore expect their research to follow their faculty, and their faculty to follow their hiring, more than their new parent unit.

\paragraph{Limitations and Future Directions}
Our descriptive analysis documents co-occurrence but cannot establish whether organizational structure shapes research, research drives structural choices, or both co-evolve.
The sample is limited to the United States, and publications leave out teaching, service, and professional impact.
Because school type is a property of the school, the tests rest on only six to 14 schools per type; they have little power, a nonsignificant difference does not show that the types are alike, and a few atypical schools can shift a type's profile.
For the same reason, we did not adjust the comparisons for the research intensity of the university or the size of the unit.
Nor did we measure the disciplinary backgrounds of the faculty, which are one likely link between a unit's organizational type and its research.
We also left out the teaching programs of the units, such as the number of degree programs they offer, even though teaching needs may drive hiring more than research does.
The changes over time combine two sources that we did not separate: faculty who changed their research and faculty who joined or left the schools.
We matched only 1{,}602 faculty members to a Dimensions profile, so the publications of the other 690 (30.1\%), more than half of them non-tenure-track, are missing.
Future research could test whether the partner discipline is linked to research through the faculty it helps to hire, through infrastructure, or through norms of evaluation.
The faculty lists of this study would also allow researchers to separate turnover from changes in the work of continuing faculty.
Other studies could follow schools through reorganizations, such as those at Wisconsin--Madison in 2019 and at Chapel Hill and Austin in 2026.
\section{Conclusion}
This study linked the organization of U.S.\ LIS schools to the research of their faculty, using 18{,}048 publications of 2013--2026 by 1{,}602 faculty members at 44 schools of five organizational types. It makes three contributions.
First, it describes LIS through the research of its faculty rather than through LIS journals. Seen this way, Human-Centered and Social Computing (36.4\% of the publications) and Health Informatics (21.1\%) are large parts of the field, and 47.2\% of faculty publications in the average school appear outside LIS venues. A landscape of LIS depends on whose research it observes.
Second, it measures how strongly organizational structure is associated with research. Faculty research profiles differ by school type, more in where faculty publish than in what they study, but type accounts for only 2.2\% of the variation among faculty members in the research themes and 5.3\% in the venue fields; most of the variation lies among the faculty of the same school. \textsc{Computer} schools lean most toward computational research and Computer Science venues, \textsc{Education} schools toward libraries, archives, and Education venues, and independent \textsc{Information} schools stay closest to the average school. Over the period, schools of every type moved away from library research and LIS venues, but seldom toward the venues of their parent unit's discipline.
Third, it releases the data: the schools and their types, the faculty lists, and the publications with their research themes and venue fields, together with LIScape, an interactive view of the corpus.

Every school type still shares the library core of LIS, but that core is shrinking in all of them. Whether this spread of LIS research across areas and venues becomes a collective strength or a loss of shared identity is a question the field will need to answer.

\section{Data Availability Statement}
The data analyzed in this study are deposited at Zenodo (\url{https://doi.org/10.5281/zenodo.18396781}). They include the 44 schools and their organizational types, the faculty lists with each matched faculty member's Dimensions researcher ID, and the 18{,}048 publications with their Dimensions publication IDs, DOIs, research themes, and venue fields. Abstracts and other Dimensions metadata are not redistributed, but each record can be retrieved from Dimensions (\url{https://app.dimensions.ai/}) by its publication ID or DOI. LIScape, an interactive view of the corpus and the theme model, is available at \url{https://jiangenhe.com/liscape/}.

\section{Uses and Roles of Generative AI}
We used generative AI for two tasks in the Methods and to improve the wording of the manuscript.
An AI browser agent (browser-use) extracted faculty names and titles from directory pages (Appendix~A), and we validated every record by hand.
Gemini 3.8 Flash proposed names for the research themes after the clustering had formed them, so it did not affect which publications belong to a theme or any statistical result; we revised eight of the 15 theme labels and all four area labels (Appendix~D).
The authors reviewed all AI-assisted output and take full responsibility for the manuscript.

\printbibliography


\end{document}


\appendix

\section{Collection of the Faculty Lists} \label{sec:prompt}
For each school and each of the four archived collection points (2014, 2017, 2020, and 2023), we searched the URL of the current faculty directory page in the Internet Archive's Wayback Machine and extracted the faculty from the archived snapshots, using Python scripts to collect the snapshot URLs, which follow the format ``https://web.archive.org/web/\{timestamp\}/\{directory\_page\_URL\}''.
Many directory URLs had changed over time, so the correct archived directory page often had to be identified by hand.
We located it from snapshots of the school's website or, where that website had been restructured, from snapshots of the university's website; in rare cases where the university's URL itself had changed, we found the historical URL with a search engine.
In this way we collected archived snapshots of the faculty directory pages of all schools for all four years.
An AI browser agent (browser-use, running OpenAI's GPT-4o at temperature 0) extracted the faculty from each page with the following prompt.

\begin{quote}
Extract Library and Information Science faculty from \{university\} (\{year\}) at: \{archive\_url\}

TASK STEPS:

1. OPEN the supplied URL. Treat this as ``page 1'' or the ``initial view.''

2. CAPTURE all faculty blocks currently visible.

3. DETECT continuation mechanisms:

    - Numbered pagination (e.g., 1 2 3, ``Next'', ``Previous'').

    - A ``Load More'', ``Show All'', or similarly-worded button.

    - Ignore A-Z or other alphabetical filters; never click them.

4. IF a continuation mechanism exists, follow it sequentially (Next → Next … or 2 → 3 → 4 …) until no new faculty appear. After each navigation, repeat step 2.

5. WHEN no further pages/loads remain, proceed to ``Finalize Output.''

FACULTY TO EXTRACT:

- Full Name (required)

- Title/Position (optional) 

- Department (optional)

- Email (optional)

INCLUDE: Professors, Associate/Assistant Professors (full-time only)

EXCLUDE: Visiting, adjunct, part-time, students, staff.
\end{quote}

\section{Organizational Type Coding of LIS Schools}
\label{sec:school_coding}
Each LIS unit was coded by its administrative parent. (1)~If the unit reports directly to the university, it is coded \textsc{Information}. (2)~If its parent is a college of arts and sciences, or a college or division of one branch of the arts and sciences such as the social sciences, it is coded \textsc{Art\&Science}; such units span too many disciplines for sibling units to indicate a partner discipline. (3)~Otherwise, the type is set by the sibling units that share the parent: computer science or another computing unit (\textsc{Computer}), communication or journalism (\textsc{Communication}), or education (\textsc{Education}); a parent with none of these siblings gives no type. Because many units were reorganized during the observation window, we traced the parent of each LIS unit from January 2013 to September 2026, when the last faculty lists were collected, in university announcements, catalogs, and archived websites. We dated each change to the month it took effect (July, the start of the academic year, when only the year was known), coded each period by the rule, and assigned each school the type it held for the longest time. Every school held one type for more than half of the 13.75-year window. Seventeen schools changed their parent unit at least once. For example, the library and information science program of the University of Hawaii--Manoa was part of the Department of Information and Computer Sciences until June 2022 and has since been part of a School of Communication and Information; it held the \textsc{Computer} type for 9.5 of the 13.75 years and is coded \textsc{Computer}. The school of the University of Wisconsin--Milwaukee, which was independent until it joined a college with the School of Education in July 2023, is coded \textsc{Information}. Table~\ref{tab:school_coding} gives each LIS unit, the parent and sibling units that determined its type, and every change of parent in the window, with the years under each type (websites accessed September and October 2026).

\begin{landscape}
\singlespacing
\footnotesize
\setlength{\tabcolsep}{4pt}
\renewcommand{\arraystretch}{1.25}
\begin{longtable}{>{\raggedright\arraybackslash}p{0.13\linewidth}>{\raggedright\arraybackslash}p{0.15\linewidth}>{\raggedright\arraybackslash}p{0.17\linewidth}>{\raggedright\arraybackslash}p{0.13\linewidth}>{\raggedright\arraybackslash}p{0.34\linewidth}}
\caption{Parent Units, Deciding Sibling Unit, and Organizational Type of Each LIS School, 2013--2026}
\label{tab:school_coding}\\
\toprule
\textbf{University} & \textbf{LIS unit} & \textbf{Parent unit} & \textbf{Deciding sibling unit} & \textbf{Years under each type and changes of parent unit, 2013--2026} \\
\midrule
\endfirsthead
\multicolumn{5}{l}{\textit{Table~\ref{tab:school_coding} (continued)}} \\
\toprule
\textbf{University} & \textbf{LIS unit} & \textbf{Parent unit} & \textbf{Deciding sibling unit} & \textbf{Years under each type and changes of parent unit, 2013--2026} \\
\midrule
\endhead
\midrule
\multicolumn{5}{r}{\textit{Continued on next page}} \\
\endfoot
\bottomrule
\multicolumn{5}{p{0.95\linewidth}}{\textit{Note.} The parent unit and deciding sibling unit are those of the period that determined the type. Years are counted from January 2013 to September 2026 (13.75 years).} \\
\endlastfoot
\multicolumn{5}{l}{\textsc{Information} (14): reports directly to the university for most of the window} \\
\midrule
Emporia State University & \lnk{https://www.emporia.edu/school-library-and-information-management/}{School of Library and Information Management} & University & --- & Unchanged (\textsc{Information}, 13.8 yr). University Libraries and Archives (2021) and Instructional Design and Technology (2023) were merged into the school, so they are internal units, not siblings. \\
North Carolina Central University & \lnk{https://www.nccu.edu/slis}{School of Library and Information Sciences} & \lnk{https://myeol.nccu.edu/sites/default/files/2025-02/University_Organization_Structure-2025.pdf}{University (organization chart)} & --- & Unchanged (\textsc{Information}, 13.8 yr). Independent school reporting to the provost. \\
Simmons University & \lnk{https://www.simmons.edu/directory/departments/slis}{School of Library and Information Science} & \lnk{https://www.simmons.edu/academics/colleges-schools-departments}{University (six-school structure)} & --- & \textsc{Information} 7.8 yr, \textsc{Computer} 6.0 yr. Independent school until June 2018; from July 2018 to June 2024 in the \lnk{https://web.archive.org/web/20180714101443/http://www.simmons.edu/academics/colleges}{College of Organizational, Computational, and Information Sciences}, with Computer Science and the School of Business; independent again since the \lnk{https://www.simmons.edu/news/welcome-fall-2024-semester}{2024 restructuring}. \\
Syracuse University & \lnk{https://ischool.syracuse.edu/}{School of Information Studies} & University & --- & Unchanged (\textsc{Information}, 13.8 yr). \\
University of Illinois--Urbana-Champaign & \lnk{https://ischool.illinois.edu/}{School of Information Sciences} & University & --- & Unchanged (\textsc{Information}, 13.8 yr). \\
University of Iowa & \lnk{https://slis.uiowa.edu/}{School of Library and Information Science} & \lnk{https://grad.uiowa.edu/}{Graduate College} & --- & Unchanged (\textsc{Information}, 13.8 yr). The school has been in the Graduate College since 2000. The Graduate College is the university-wide graduate administration, not a disciplinary college, so the school is treated as reporting to the university. \\
University of Maryland--College Park & \lnk{https://ischool.umd.edu/}{College of Information} & University & --- & Unchanged (\textsc{Information}, 13.8 yr). \\
University of Michigan--Ann Arbor & \lnk{https://www.si.umich.edu/}{School of Information} & University & --- & Unchanged (\textsc{Information}, 13.8 yr). \\
University of North Carolina--Chapel Hill & \lnk{https://sils.unc.edu/}{School of Information and Library Science} & University & --- & \textsc{Information} 13.5 yr, \textsc{Computer} 0.3 yr. Independent school until June 2026; since July 2026 \lnk{https://www.unc.edu/posts/2026/07/01/carolinas-school-of-data-and-information-sciences-launches/}{merged with the School of Data Science and Society} into the School of Data and Information Sciences. \\
University of North Texas & \lnk{https://ci.unt.edu/}{College of Information} & University & --- & Unchanged (\textsc{Information}, 13.8 yr). The iSchool is the College of Information itself; its departments (\lnk{https://informationscience.unt.edu/}{Information Science}, Learning Technologies, Data Science) are internal units. \\
University of Texas--Austin & \lnk{https://ischool.utexas.edu/}{School of Information} & University & --- & \textsc{Information} 13.7 yr, \textsc{Computer} 0.1 yr. Independent school until August 2026; since fall 2026 a founding department of the \lnk{https://news.utexas.edu/2026/02/19/ut-launches-new-school-of-computing-uniting-computer-and-data-science-statistics-information-disciplines/}{School of Computing} in the College of Natural Sciences. \\
University of Washington & \lnk{https://ischool.uw.edu/}{Information School} & University & --- & Unchanged (\textsc{Information}, 13.8 yr). \\
University of Wisconsin--Milwaukee & \lnk{https://uwm.edu/informationstudies/}{School of Information Studies} & University & --- & \textsc{Information} 10.5 yr, \textsc{Education} 3.3 yr. Independent school until June 2023; since July 2023 in the \lnk{https://uwm.edu/community-engagement-professions/}{College of Community Engagement \& Professions}, with the \lnk{https://uwm.edu/education/}{School of Education} (\lnk{https://uwm.edu/news/uwms-realignment-of-schools-and-colleges-now-in-effect/}{realignment of July 2023}). \\
Wayne State University & \lnk{https://sis.wayne.edu/}{School of Information Sciences} & \lnk{https://bulletins.wayne.edu/undergraduate/school-information-sciences/}{University (bulletin)} & --- & Unchanged (\textsc{Information}, 13.8 yr). Independent school. \\
\midrule
\multicolumn{5}{l}{\textsc{Computer} (8): sibling unit is computer science or another computing unit for most of the window} \\
\midrule
Drexel University & \lnk{https://drexel.edu/cci/academics/information-science-department/}{Department of Information Science} & \lnk{https://drexel.edu/cci/}{College of Computing \& Informatics} & \lnk{https://drexel.edu/cci/academics/computer-science-department/}{Department of Computer Science} & \textsc{Computer} 13.1 yr, \textsc{Information} 0.7 yr. The independent College of Information Science and Technology merged with the Department of Computer Science and the Computing and Security Technology program into the College of Computing \& Informatics in \lnk{https://drexel.edu/news/archive/2013/september/cci-launch}{September 2013}. \\
Indiana University--Bloomington & \lnk{https://luddy.iu.edu/academics/units/ils/index.html}{Department of Information and Library Science} & \lnk{https://luddy.iu.edu/}{Luddy School of Informatics, Computing, and Engineering} & \lnk{https://luddy.iu.edu/academics/units/computer-science.html}{Computer Science} & \textsc{Computer} 13.3 yr, \textsc{Information} 0.5 yr. The independent School of Library and Information Science \lnk{https://asist-archive.ischool.illinois.edu/asis-l/2012-October/006773.html}{merged into the School of Informatics and Computing} (now the Luddy School) in July 2013. \\
Indiana University-Purdue University--Indianapolis & \lnk{https://luddy.indianapolis.iu.edu/academics/departments/lis/index.html}{Department of Library and Information Science} & \lnk{https://luddy.indianapolis.iu.edu/}{Luddy School of Informatics, Computing, and Engineering} & \lnk{https://luddy.indianapolis.iu.edu/academics/departments/hcc/index.html}{Department of Human-Centered Computing} & \textsc{Computer} 13.3 yr, \textsc{Information} 0.5 yr. Part of the independent Indiana University School of Library and Information Science until June 2013; in the School of Informatics and Computing (now the Luddy School) since the \lnk{https://asist-archive.ischool.illinois.edu/asis-l/2012-October/006773.html}{merger of July 2013}. \\
San Jose State University & \lnk{https://ischool.sjsu.edu/}{School of Information} & \lnk{https://web.archive.org/web/20191003043012/http://www.sjsu.edu/cies}{College of Professional and Global Education} (now \lnk{https://www.sjsu.edu/information-data-society/}{College of Information, Data and Society}) & \lnk{https://www.sjsu.edu/applied-data-science/}{Department of Applied Data Science} & \textsc{Computer} 7.3 yr, \textsc{Communication} 6.5 yr. Until June 2019 in the \lnk{https://web.archive.org/web/20170602211036/http://ischool.sjsu.edu/}{College of Applied Sciences and Arts} (renamed College of Health and Human Sciences in 2018), whose units included the School of Journalism and Mass Communications; since July 2019 in the College of Professional and Global Education (renamed College of Information, Data and Society in 2024), whose only other academic department is Applied Data Science. \\
University at Albany--SUNY & \lnk{https://www.albany.edu/cehc/information-sciences-and-technology}{Department of Information Sciences and Technology} & \lnk{https://www.albany.edu/cehc}{College of Emergency Preparedness, Homeland Security and Cybersecurity} & \lnk{https://www.albany.edu/cehc/cybersecurity}{Department of Cybersecurity} & \textsc{Computer} 13.8 yr, under three parents: the College of Computing and Information, with Computer Science, until June 2015; the \lnk{https://web.archive.org/web/20170528185215/http://www.albany.edu/informationstudies/}{College of Engineering and Applied Sciences}, with Computer Science, from July 2015; and the College of Emergency Preparedness, Homeland Security and Cybersecurity since \lnk{https://www.albany.edu/news/84098.php}{January 2018}. \\
University of Hawaii--Manoa & \lnk{https://sci.manoa.hawaii.edu/programs/library-and-information-science/}{Library and Information Science Program} & \lnk{https://www.ics.hawaii.edu/}{Department of Information and Computer Sciences} (College of Natural Sciences) & \lnk{https://www.ics.hawaii.edu/}{Computer science programs of the department} & \textsc{Computer} 9.5 yr, \textsc{Communication} 4.3 yr. A program of the Department of Information and Computer Sciences from 1997 to June 2022; since \lnk{https://www.hawaii.edu/lis/about-us/program-history/}{July 2022} in the \lnk{https://sci.manoa.hawaii.edu/}{School of Communication and Information} (College of Social Sciences), with Communication, Communicology, and Journalism. \\
University of Pittsburgh & \lnk{https://www.icds.pitt.edu/}{Department of Information Culture and Data Stewardship} & \lnk{https://www.sci.pitt.edu/}{School of Computing and Information} & \lnk{https://www.cs.pitt.edu/}{Department of Computer Science} & \textsc{Computer} 9.3 yr, \textsc{Information} 4.5 yr. The independent School of Information Sciences \lnk{https://www.cs.pitt.edu/about}{joined with the Department of Computer Science} to form the School of Computing and Information in July 2017. \\
University of Wisconsin--Madison & \lnk{https://ischool.wisc.edu/}{Information School} & \lnk{https://news.wisc.edu/new-uw-madison-school-sets-emphasis-on-computing-data-fields/}{School of Computer, Data \& Information Sciences} (now \lnk{https://cai.wisc.edu/}{College of Computing \& AI}) & \lnk{https://cs.wisc.edu/}{Department of Computer Sciences} & \textsc{Computer} 7.1 yr, \textsc{Art\&Science} 6.7 yr. A school of the College of Letters \& Science until August 2019; since \lnk{https://news.wisc.edu/new-uw-madison-school-sets-emphasis-on-computing-data-fields/}{September 2019} in the School of Computer, Data \& Information Sciences, with Computer Sciences and Statistics, which became the \lnk{https://cai.wisc.edu/}{College of Computing \& Artificial Intelligence} in July 2026. From 2019 to 2026 the School of Computer, Data \& Information Sciences was itself part of the College of Letters \& Science; the immediate parent is used. \\
\midrule
\multicolumn{5}{l}{\textsc{Communication} (7): sibling unit is communication or journalism for most of the window} \\
\midrule
Florida State University & \lnk{https://ischool.cci.fsu.edu/}{School of Information} & \lnk{https://cci.fsu.edu/}{College of Communication and Information} & \lnk{https://comm.cci.fsu.edu/}{School of Communication} & Unchanged (\textsc{Communication}, 13.8 yr). \\
Kent State University & \lnk{https://www.kent.edu/ischool}{School of Information} & \lnk{http://web.archive.org/web/20241211002406/https://www.kent.edu/cci}{College of Communication and Information (Dec.\ 2024 archive)} & \lnk{https://www.kent.edu/mdj}{School of Media and Journalism} & \textsc{Communication} 13.5 yr, \textsc{Art\&Science} 0.3 yr. In the College of Communication and Information until June 2026; since July 2026 in the \lnk{https://www.kent.edu/sh/news/one-college-many-disciplines-message-dean}{College of Sciences and Humanities} (the former College of Arts and Sciences), under a restructuring approved in 2025. \\
Rutgers University--New Brunswick & \lnk{https://sci.rutgers.edu/departments/library-and-information-science-department}{Department of Library and Information Science} & \lnk{https://sci.rutgers.edu/}{School of Communication and Information} & \lnk{https://sci.rutgers.edu/departments/communication-department}{Department of Communication} & Unchanged (\textsc{Communication}, 13.8 yr). \\
University of Alabama & \lnk{https://cis.ua.edu/departments/slis/}{School of Library and Information Studies} & \lnk{https://cis.ua.edu/}{College of Communication and Information Sciences} & \lnk{https://cis.ua.edu/departments/communication-studies/}{Department of Communication Studies} & Unchanged (\textsc{Communication}, 13.8 yr). \\
University of Kentucky & \lnk{https://ci.uky.edu/sis/}{School of Information Science} & \lnk{https://ci.uky.edu/}{College of Communication and Information} & \lnk{https://comm.uky.edu/}{Department of Communication} & Unchanged (\textsc{Communication}, 13.8 yr). \\
University of South Carolina & \lnk{https://sc.edu/study/colleges_schools/cic/library_and_information_science/index.php}{School of Information Science} & \lnk{https://sc.edu/study/colleges_schools/cic/index.php}{College of Information and Communications} & \lnk{https://sc.edu/study/colleges_schools/cic/journalism_and_mass_communications/index.php}{School of Journalism and Mass Communications} & Unchanged (\textsc{Communication}, 13.8 yr). \\
University of Tennessee--Knoxville & \lnk{https://cci.utk.edu/sis}{School of Information Sciences} & \lnk{https://cci.utk.edu/}{College of Communication and Information} & \lnk{https://cci.utk.edu/cmst}{School of Communication Studies} & Unchanged (\textsc{Communication}, 13.8 yr). \\
\midrule
\multicolumn{5}{l}{\textsc{Education} (9): sibling unit is education for most of the window} \\
\midrule
Long Island University Post & \lnk{https://www.liu.edu/palmer/}{Palmer School of Library and Information Science} & \lnk{https://liu.edu/post/ceit}{College of Education, Information and Technology} & \lnk{https://www.liu.edu/post/academics/school-of-education}{School of Education} & Unchanged (\textsc{Education}, 13.8 yr). \\
Louisiana State University--Baton Rouge & \lnk{https://www.lsu.edu/chse/sis/}{School of Information Studies} & \lnk{https://www.lsu.edu/chse/}{College of Human Sciences \& Education} & \lnk{https://www.lsu.edu/chse/education/}{School of Education} & Unchanged (\textsc{Education}, 13.8 yr); the college was formed in 2012. \\
Texas Woman's University & \lnk{https://twu.edu/slis/}{School of Library and Information Studies} & \lnk{https://twu.edu/college-professional-education/}{College of Professional Education} & \lnk{https://twu.edu/education/}{School of Education} & Unchanged (\textsc{Education}, 13.8 yr). \\
University at Buffalo--SUNY & \lnk{https://ed.buffalo.edu/information.html}{Department of Information Science} & \lnk{https://ed.buffalo.edu/}{Graduate School of Education} & \lnk{https://ed.buffalo.edu/teaching.html}{Department of Learning and Instruction} & Unchanged (\textsc{Education}, 13.8 yr). \\
University of California--Los Angeles & \lnk{https://seis.ucla.edu/academics/department-of-information-studies/}{Department of Information Studies} & \lnk{https://seis.ucla.edu/}{School of Education and Information Studies} & \lnk{https://seis.ucla.edu/academics/department-of-education/}{Department of Education} & Unchanged (\textsc{Education}, 13.8 yr). \\
University of Denver & \lnk{https://morgridge.du.edu/academic-programs/library-information-science}{Library and Information Science} & \lnk{https://morgridge.du.edu/}{Morgridge College of Education} & \lnk{https://morgridge.du.edu/academic-programs/curriculum-instruction}{Curriculum and Instruction} & Unchanged (\textsc{Education}, 13.8 yr). \\
University of Missouri & \lnk{https://cehd.missouri.edu/information-science-learning-technologies/}{School of Information Science \& Learning Technologies} & \lnk{https://cehd.missouri.edu/}{College of Education \& Human Development} & \lnk{https://cehd.missouri.edu/learning-teaching-curriculum/}{Learning, Teaching \& Curriculum} & Unchanged (\textsc{Education}, 13.8 yr). \\
University of North Carolina at Greensboro & \lnk{https://soe.uncg.edu/ilrs/}{Department of Information, Library, and Research Sciences} & \lnk{https://soe.uncg.edu/}{School of Education} & \lnk{https://soe.uncg.edu/tehe/}{Teacher Education and Higher Education} & Unchanged (\textsc{Education}, 13.8 yr). The Department of Library and Information Science merged into this department in 2023. \\
University of Southern Mississippi & \lnk{https://www.usm.edu/library-information-science/index.php}{School of Library and Information Science} & \lnk{https://www.usm.edu/education-human-sciences/index.php}{College of Education and Human Sciences} & \lnk{https://www.usm.edu/education/index.php}{School of Education} & \textsc{Education} 13.8 yr, under two parents: the \lnk{https://web.archive.org/web/20130520190628/http://www.usm.edu/education-psychology}{College of Education and Psychology} until June 2018, and since July 2018 the College of Education and Human Sciences, formed from it and the College of Health. \\
\midrule
\multicolumn{5}{l}{\textsc{Art\&Science} (6): parent unit is a college or division of arts and sciences for most of the window} \\
\midrule
CUNY--Queens College & \lnk{https://www.qc.cuny.edu/academics/sis/}{Graduate School of Library and Information Studies} & \lnk{https://www.qc.cuny.edu/academics/ss/}{School of Social Sciences} & --- & Unchanged (\textsc{Art\&Science}, 13.8 yr). The parent is the social sciences division of the liberal arts and sciences, and its other units are arts and sciences disciplines (e.g., \lnk{https://www.qc.cuny.edu/academics/sociology}{Sociology}, History, Anthropology). \\
The Catholic University of America & \lnk{https://lis.catholic.edu/}{Department of Information Sciences (formerly Library and Information Science)} & \lnk{https://catholic.smartcatalogiq.com/en/2024-2025/graduate-announcements/school-of-arts-and-sciences/departments-and-programs-in-the-school-of-arts-and-sciences/department-of-information-sciences/}{School of Arts and Sciences} & --- & \textsc{Art\&Science} 13.3 yr, \textsc{Information} 0.5 yr. The independent School of Library and Information Science \lnk{https://lis.catholic.edu/lisadvisoryboard-meeting-2014fall-presentation-2014-11-20.pdf}{became a department} of the School of Arts and Sciences in 2013. \\
Dominican University & \lnk{https://www.dom.edu/admission/graduate/college-applied-social-sciences/library-and-information-studies-programs/school}{School of Information Studies} & \lnk{https://web.archive.org/web/20200531084614/https://www.dom.edu/academics}{College of Applied Social Sciences (2020 archive)} & --- & \textsc{Art\&Science} 8.0 yr, \textsc{Information} 4.5 yr, no type 1.3 yr. Independent school until June 2017; from July 2017 to June 2025 in the College of Applied Social Sciences, a college of the applied social sciences with the School of Education and the School of Social Work (\lnk{https://dom.catalog.acalog.com/index.php?catoid=10}{2017--2018 bulletin}); since July 2025 in the \lnk{https://www.oakpark.com/2025/07/08/dominican-realigns-around-tech-health-care-career-training/}{College of Business, Information Studies, and Technology}, with the Brennan School of Business. \\
University of Arizona & \lnk{https://infosci.arizona.edu/}{College of Information Science} & \lnk{https://web.archive.org/web/20190415024558/https://ischool.arizona.edu/}{College of Social and Behavioral Sciences (2019 archive)} & --- & \textsc{Art\&Science} 10.5 yr, \textsc{Information} 3.3 yr. In the College of Social and Behavioral Sciences until June 2023, as the School of Information Resources and Library Science and, from 2015, as the School of Information, formed with a school of the College of Science; an \lnk{https://registrar.arizona.edu/sites/default/files/iSchool\%20program\%20\%26\%20org_Approval\%20Memo\%20\%26\%20Details.pdf}{independent college} since \lnk{https://infosci.arizona.edu/news/u-launches-standalone-ischool-first-state}{July 2023}, renamed College of Information Science in 2024. \\
University of Oklahoma & \lnk{https://www.ou.edu/cas/slis}{School of Library and Information Studies} & \lnk{https://www.ou.edu/cas}{Dodge Family College of Arts and Sciences} & --- & Unchanged (\textsc{Art\&Science}, 13.8 yr). \\
University of South Florida & \lnk{https://www.usf.edu/arts-sciences/departments/information/}{School of Information} & \lnk{https://www.usf.edu/arts-sciences/}{College of Arts and Sciences} & --- & Unchanged (\textsc{Art\&Science}, 13.8 yr). \\
\end{longtable}
\end{landscape}

\section{Matching Faculty Members and Selecting Publications}
\label{sec:pubfilter}
\textit{Matching faculty members to Dimensions.} We used the Dimensions Analytics API (DSL v2). For each faculty member, we first issued an exact-name query constrained by institutional affiliation, \textit{search researchers where first\_name = "\{firstname\}" and last\_name = "\{lastname\}" and research\_orgs.id = "\{grid\_id\}" return researchers}, where \{grid\_id\} is the GRID identifier of the faculty member's university. If the exact query returned no result, we relaxed the first-name constraint to a fuzzy match, \textit{first\_name $\sim$ "\{firstname\}"}, while keeping the affiliation filter. When several researcher records were returned for a faculty member, we retrieved recent publications for each candidate and identified by hand those working in Library and Information Science or closely related areas. We then verified every researcher ID against its Dimensions record (name, affiliations, and publication titles). Dimensions sometimes splits one researcher into several profiles, or retires a profile after merging it into another; we therefore added the other profiles of the same person, confirmed by shared co-authors or publication topics, so that 154 faculty members are represented by more than one profile. We retrieved the publications of each profile with \textit{search publications where researchers = "\{dimension\_id\}" return publications}.

\textit{Selection of publications.} Because publishing norms differ across the areas of LIS, we kept journal articles, conference papers, books, and book chapters, including records to which Dimensions assigns no document type, and removed records that do not report full research or that have not been peer reviewed. We applied four criteria in turn to the 26{,}053 unique publications and counted each removed record under the first criterion it meets. Figure~2 in the main text shows the whole selection. The 18{,}122 publications that pass the selection include 1{,}606 without a document type in Dimensions, among them all 130 books.

\textit{Document type.} We removed 931 records of Dimensions document types that do not report full research: editorials (315), conference abstracts (294), book reviews (118), corrections (64), letters to the editor (56), and the front and back matter of books and proceedings, such as prefaces, chairs' messages, keynote listings, and tutorials (84). We kept the type for other journal content, which holds research articles as well as columns; the third criterion removes the front matter among them.

\textit{Preprints.} We removed the 3{,}508 records of the Dimensions type preprint, which have not been peer reviewed. Of these, 1{,}620 (46\%) have the same title, ignoring case and punctuation, as a publication we retained, so many of the studies they report enter the corpus in their published version.

\textit{Front and back matter identified by title.} We removed 110 records whose title is only the name of a front- or back-matter section, such as ``Introduction,'' ``Preface,'' ``From the Editor's Desk,'' ``President's Page,'' ``Message from the Program Chairs,'' ``In Memoriam,'' or ``Session details,'' which Dimensions leaves without a document type or labels as articles or chapters.

\textit{Short conference contributions.} We removed 3{,}382 conference contributions that meet at least one of three criteria: papers in the Extended Abstracts, Companion, and Adjunct volumes of conferences, which hold late-breaking work, posters, workshop papers, and demonstrations (647); conference papers of at most six pages (3{,}127); and papers whose title identifies them as a poster, demonstration, panel, workshop proposal or report, or doctoral-consortium paper (199). Of the 3{,}382, 2{,}556 meet only the length criterion, 185 only the volume criterion, and 64 only the title criterion. The title criterion spares research papers whose titles use these words for their subject, such as panel studies or studies of design workshops.
We counted as conference papers those that Dimensions types as proceedings papers and those in proceedings that Dimensions files as journals or book series, such as the Proceedings of the Association for Information Science and Technology and the Proceedings of the ALISE Annual Conference, listed in the main text as the ASIS\&T Annual Meeting and the ALISE Annual Conference, and Lecture Notes in Computer Science. Journals named for proceedings, such as the Proceedings of the National Academy of Sciences and the Proceedings of the ACM on Human-Computer Interaction, were treated as journals. The length comes from the page range; papers without one, such as those numbered by article, were kept. Journal articles were kept whatever their length.

\section{Choosing the Clustering Method and the Number of Research Themes}
\label{sec:ksweep}
We clustered the publications with agglomerative clustering and Ward linkage rather than HDBSCAN (Hierarchical Density-Based Spatial Clustering of Applications with Noise). We tried both, and Ward linkage gave more balanced and more interpretable themes for our corpus.

Ward linkage is deterministic and hierarchical, so every candidate solution is a cut of one linkage over the ten-dimensional UMAP coordinates of the 15{,}934 SPECTER2 embeddings; the 15 themes of the main text are the $K=15$ cut of this linkage. We cut it at $K \in \{3, 4, 5, 6, 8, 10, 12, \ldots, 18, 20, 22, 24, 25, 26, 28, \ldots, 40\}$, that is, at every value from 12 to 18, at every even value from 20 to 40, and at 25. The cuts are nested, so a theme at a finer cut is part of a theme at every coarser one.
Term-based scores use one vocabulary for all cuts: the 9{,}787 terms that occur in at least ten title--abstract documents, after removing English stop words.
We scored each cut on five measures.

\textit{Coherence} is the mean pairwise normalized pointwise mutual information \parencite[NPMI;][]{bouma2009normalized} of each theme's ten top c-TF-IDF terms over the binary document--term matrix, with never co-occurring pairs contributing $-1$.
\textit{Term diversity} is the share of distinct terms among the ten top terms of all themes \parencite{dieng2020topic}.
\textit{Fields of Research agreement} is external to our text representation. Dimensions assigns each publication to groups of the ANZSRC 2020 Fields of Research classification; we use the 165 four-digit groups, which cover 95.1\% of the publications. For each theme we take the mean Jaccard similarity between the group sets of all pairs of its publications, average this over themes, and divide by the mean over 200{,}000 random pairs of publications (0.097).
\textit{Subsample stability} is the mean adjusted Rand index \parencite[ARI;][]{hubert1985comparing} between a cut of the full corpus and the same cut of a Ward linkage re-run on an 80\% subsample, over ten subsamples, computed on the sampled publications.
\textit{Split-half stability} re-runs the whole pipeline, UMAP and Ward, on two disjoint random halves of the corpus, each with its own random seed \parencite{lange2004stability}. Because the halves share no publications, we carry one half's themes onto the other: each publication in one half receives the majority theme of its 15 nearest neighbors (cosine similarity of SPECTER2 embeddings) in the other half. We report the ARI between each half's own themes and the themes carried from the other half, over ten random splits in both directions (20 comparisons).

\begin{figure}[!htbp]
    \centering
    \includegraphics[width=\linewidth]{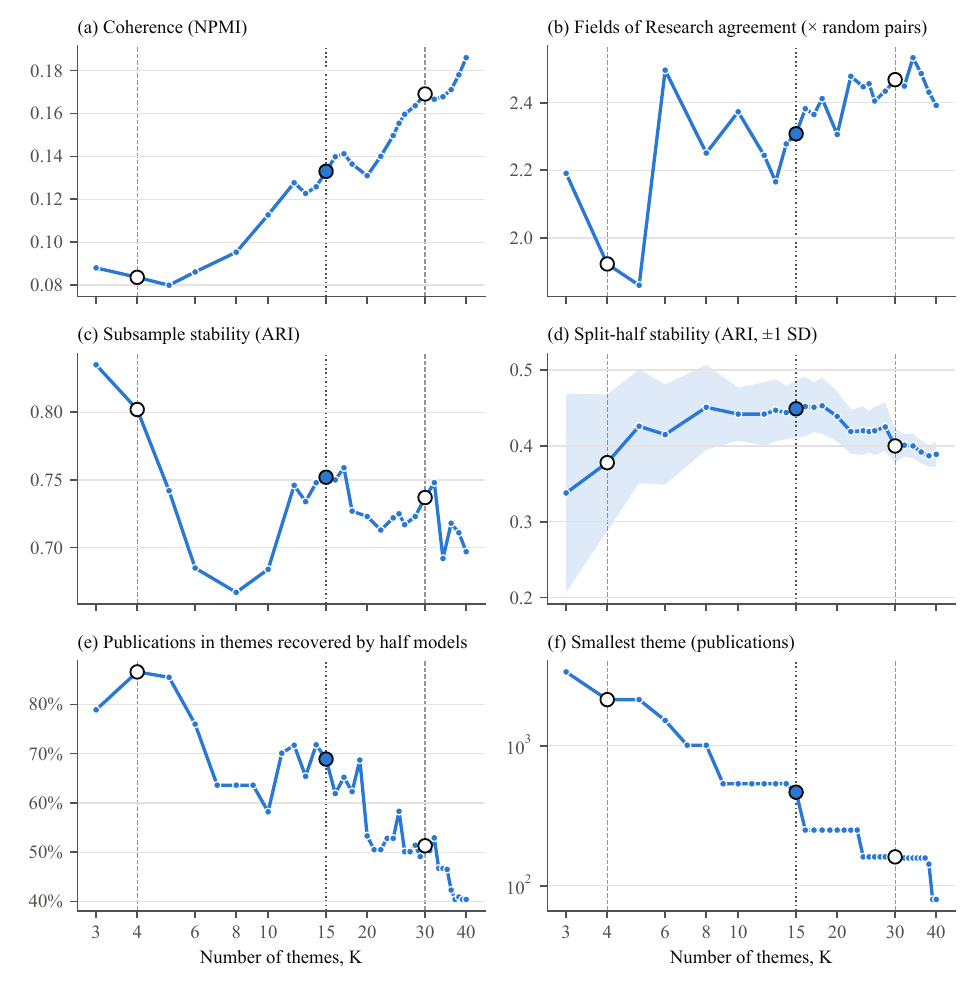}
    \caption{Choosing the number of research themes. Each point is a cut of the same Ward linkage; the dotted line and the filled marker show the themes of the main text ($K=15$), and the dashed lines and open markers the coarse ($K=4$) and fine ($K=30$) levels. The shaded band in panel (d) is one standard deviation over 20 split-half comparisons. Panels (e) and (f) cover every cut from $K=3$ to $K=40$: (e) is the share of publications in themes recovered in at least half of the 20 half models (Section~\ref{sec:k14replication}), and (f) is the size of the smallest theme, on a logarithmic scale.}
    \label{fig:ksweep}
\end{figure}

Table~\ref{tab:ksweep} and Figure~\ref{fig:ksweep} report the sweep.
Term diversity stays within a narrow band (0.64--0.73) and does not separate the solutions.
Coherence is lowest for the coarsest solutions (0.080--0.095 up to $K=8$) and rises with the number of themes, to 0.128 at $K=12$, 0.133 at $K=15$, and 0.186 at $K=40$. Fields of Research agreement varies between 1.86 and 2.53 without a clear trend; it is lowest at $K=4$ and $K=5$ (1.92 and 1.86) and at least 2.28 from $K=14$ on.
Subsample stability varies between 0.67 and 0.84; it is highest for the coarsest solutions (0.84 at $K=3$ and 0.80 at $K=4$), lowest from $K=6$ to $K=10$ (0.67--0.69), and between 0.69 and 0.76 from $K=12$ on. Split-half stability is lowest for the coarsest solutions: it rises from 0.338 at $K=3$ to 0.426 at $K=5$, stays between 0.442 and 0.453 from $K=8$ to $K=18$, with its maximum at $K=18$, and declines from $K=20$ on, to 0.419--0.425 from $K=22$ to $K=28$ and 0.387--0.401 from $K=30$ to $K=40$. The coarse solutions also vary the most from one split to the next (standard deviation 0.131 at $K=3$ and 0.090 at $K=4$, against 0.014--0.042 for $K \ge 10$).

The measures therefore identify a range rather than a single number. From $K=13$ to $K=18$, split-half stability stays within 0.01 of its maximum (0.444--0.453), less than one standard deviation over splits, and coherence and subsample stability are higher than for the solutions from $K=6$ to $K=10$; $K=8$ is as stable across split halves (0.451) but far less coherent (0.095). Within this range, $K=14$ and $K=15$ hold the largest shares of publications in themes that are recovered in at least half of the half models (Sections~\ref{sec:k14replication} and~\ref{sec:klevels}): 72\% and 69\% (Figure~\ref{fig:ksweep}e). We retained $K=15$. It differs from $K=14$ by one split, which separates Scientometrics and Science of Science (774 publications), on citation analysis, scientific careers, and research integrity, from Data Curation and Cyberinfrastructure (466), on the provenance and curation of scientific data, citizen science, and research workflows; at $K=14$ the two form one theme that is recovered in 60\% of the half models. We kept them apart because citation analysis and the curation of scientific data are distinct research areas, and the split leaves Scientometrics and Science of Science as one of the most reproducible themes (recovered in 90\% of the half models), although Data Curation and Cyberinfrastructure is recovered by no half model. We did not go further: the split that $K=16$ adds divides Cybersecurity and Information Privacy, which is recovered in 95\% of the half models, into two parts recovered in 10\% and 45\%, and lowers the share of publications in recovered themes to 62\%.

\textit{Number of UMAP dimensions.} The ten dimensions of the UMAP projection are also a choice. We re-ran UMAP with 5, 10, 15, 20, 30, 40, and 50 dimensions, with five random seeds each, cut the Ward linkage of each run at $K=15$, and compared the result with the 15 themes. With ten dimensions and other seeds, the ARI with the 15 themes averaged 0.73 (0.68--0.76); with the other numbers of dimensions it averaged between 0.70 (30 dimensions) and 0.75 (40 dimensions), and ranged from 0.65 to 0.80 over the 30 runs. Coherence (NPMI 0.132--0.137 on average) and Fields of Research agreement (2.30--2.36) hardly changed with the number of dimensions. The number of dimensions thus changes the themes no more than the random seed of UMAP does. The themes that these runs recover least often, in 26\% to 71\% of the 34 runs other than ours, are the four that the half models recover least (Table~\ref{tab:k14replication}): Accessibility and Extended Reality, Crisis Informatics and Urban Computing, Data Curation and Cyberinfrastructure, and Human-AI Interaction and Visualization.

\begin{table}[!htbp]
    \centering
    \caption{Scores of the Candidate Numbers of Research Themes}
    \label{tab:ksweep}
    \footnotesize
    \begin{tabular}{rrrrrrr}
    \toprule
    $K$ & Coherence & Diversity & FoR & Subsample & Split-half & Smallest \\
        & (NPMI)    &           & agreement & stability (ARI) & stability (ARI, SD) & theme \\
    \midrule
    3 & 0.088 & 0.733 & 2.19 & 0.835 & 0.338 (0.131) & 3{,}362 \\
    4 & 0.084 & 0.700 & 1.92 & 0.802 & 0.378 (0.090) & 2{,}132 \\
    5 & 0.080 & 0.700 & 1.86 & 0.742 & 0.426 (0.075) & 2{,}132 \\
    6 & 0.086 & 0.683 & 2.50 & 0.685 & 0.415 (0.066) & 1{,}514 \\
    8 & 0.095 & 0.688 & 2.25 & 0.667 & 0.451 (0.056) & 1{,}008 \\
    10 & 0.113 & 0.680 & 2.37 & 0.684 & 0.442 (0.035) & 535 \\
    12 & 0.128 & 0.658 & 2.24 & 0.746 & 0.442 (0.042) & 535 \\
    13 & 0.123 & 0.638 & 2.17 & 0.734 & 0.447 (0.041) & 535 \\
    14 & 0.126 & 0.679 & 2.28 & 0.748 & 0.444 (0.035) & 535 \\
    \textbf{15} & \textbf{0.133} & \textbf{0.673} & \textbf{2.31} & \textbf{0.752} & \textbf{0.449 (0.038)} & \textbf{466} \\
    16 & 0.140 & 0.681 & 2.38 & 0.750 & 0.452 (0.039) & 250 \\
    17 & 0.141 & 0.682 & 2.37 & 0.759 & 0.451 (0.033) & 250 \\
    18 & 0.136 & 0.656 & 2.41 & 0.727 & 0.453 (0.037) & 250 \\
    20 & 0.131 & 0.640 & 2.31 & 0.723 & 0.439 (0.033) & 250 \\
    22 & 0.140 & 0.636 & 2.48 & 0.713 & 0.419 (0.029) & 250 \\
    24 & 0.150 & 0.654 & 2.45 & 0.722 & 0.420 (0.032) & 161 \\
    25 & 0.155 & 0.648 & 2.46 & 0.725 & 0.419 (0.027) & 161 \\
    26 & 0.160 & 0.650 & 2.40 & 0.717 & 0.420 (0.032) & 161 \\
    28 & 0.164 & 0.657 & 2.43 & 0.723 & 0.425 (0.032) & 161 \\
    30 & 0.169 & 0.660 & 2.47 & 0.737 & 0.400 (0.022) & 161 \\
    32 & 0.167 & 0.656 & 2.45 & 0.748 & 0.401 (0.015) & 158 \\
    34 & 0.168 & 0.647 & 2.53 & 0.692 & 0.400 (0.016) & 158 \\
    36 & 0.171 & 0.647 & 2.49 & 0.718 & 0.392 (0.015) & 158 \\
    38 & 0.178 & 0.650 & 2.43 & 0.711 & 0.387 (0.014) & 143 \\
    40 & 0.186 & 0.655 & 2.39 & 0.697 & 0.389 (0.017) & 80 \\
    \bottomrule
    \end{tabular}

    \vspace{4pt}
    \begin{minipage}{0.95\linewidth}
    \footnotesize\textit{Note.} Each row is a cut of the same Ward linkage; the $K=15$ row is the theme model of the main text. FoR agreement is the within-theme Jaccard similarity of Fields of Research groups divided by that of random pairs. Split-half stability is the mean over 20 comparisons, with its standard deviation in parentheses.
    \end{minipage}
\end{table}

\subsection{Recovery of the 15 Themes by Half-Corpus Models}
\label{sec:k14replication}
We compared each 15-theme model fitted on half of the corpus with the 15 themes of the same publications. Over the 20 half models, agreement was ARI $= 0.52$ (SD 0.03).
For each theme, we took its best Jaccard match among the themes of each half model \parencite{hennig2007cluster} and counted the theme as recovered when that match is at least 0.5; by construction, no two themes can both exceed 0.5 with the same half-model theme.

Table~\ref{tab:k14replication} reports the result. Six of the 15 themes are recovered in at least 90\% of the 20 half models, and Libraries and Librarianship, Biomedical Informatics, and Human-Centered Health Informatics in all of them; nine are recovered in at least half of them. Three themes are recovered in none of the half models: Human-AI Interaction and Visualization, Accessibility and Extended Reality, and Data Curation and Cyberinfrastructure, and Crisis Informatics and Urban Computing is recovered in 15\%. The boundaries of these themes move from one run to the next.

\begin{table}[!htbp]
    \centering
    \caption{Recovery of the 15 Themes by Models Fitted on Half of the Corpus}
    \label{tab:k14replication}
    \footnotesize
    \begin{tabular}{lrrrr}
    \toprule
    \textbf{Theme} & \textbf{Publications} & \textbf{Mean} & \textbf{Min.} & \textbf{Recovered} \\
     &  & \textbf{Jaccard} & \textbf{Jaccard} & \textbf{(\%)} \\
    \midrule
    Libraries and Librarianship & 1{,}514 & 0.75 & 0.60 & 100 \\
    Biomedical Informatics & 1{,}008 & 0.74 & 0.63 & 100 \\
    Human-Centered Health Informatics & 1{,}322 & 0.68 & 0.57 & 100 \\
    Scientometrics and Science of Science & 774 & 0.68 & 0.44 & 90 \\
    Cybersecurity and Information Privacy & 1{,}121 & 0.63 & 0.48 & 95 \\
    Computing Education and Learning Technologies & 1{,}032 & 0.58 & 0.38 & 80 \\
    Social Media Information Dynamics & 1{,}283 & 0.54 & 0.43 & 90 \\
    Archives, Metadata, and Data Sharing & 1{,}893 & 0.52 & 0.42 & 60 \\
    Information Seeking and Retrieval & 535 & 0.50 & 0.30 & 40 \\
    Digital Health Information Behavior & 1{,}032 & 0.48 & 0.23 & 50 \\
    Sociotechnical and Critical Informatics & 1{,}030 & 0.44 & 0.21 & 40 \\
    Crisis Informatics and Urban Computing & 1{,}011 & 0.38 & 0.19 & 15 \\
    Accessibility and Extended Reality & 583 & 0.34 & 0.13 & 0 \\
    Human-AI Interaction and Visualization & 1{,}330 & 0.28 & 0.17 & 0 \\
    Data Curation and Cyberinfrastructure & 466 & 0.23 & 0.10 & 0 \\
    \bottomrule
    \end{tabular}

    \vspace{4pt}
    \begin{minipage}{0.95\linewidth}
    \footnotesize\textit{Note.} Jaccard is the overlap between a theme and its best-matching theme in a half model, computed on the publications of that half. Mean and minimum are over the 20 half models; a theme is recovered when the match is at least 0.5.
    \end{minipage}
\end{table}

\subsection{The Themes at a Coarser and a Finer Level}
\label{sec:klevels}
\begin{figure}[!htbp]
    \centering
    \includegraphics[width=\linewidth]{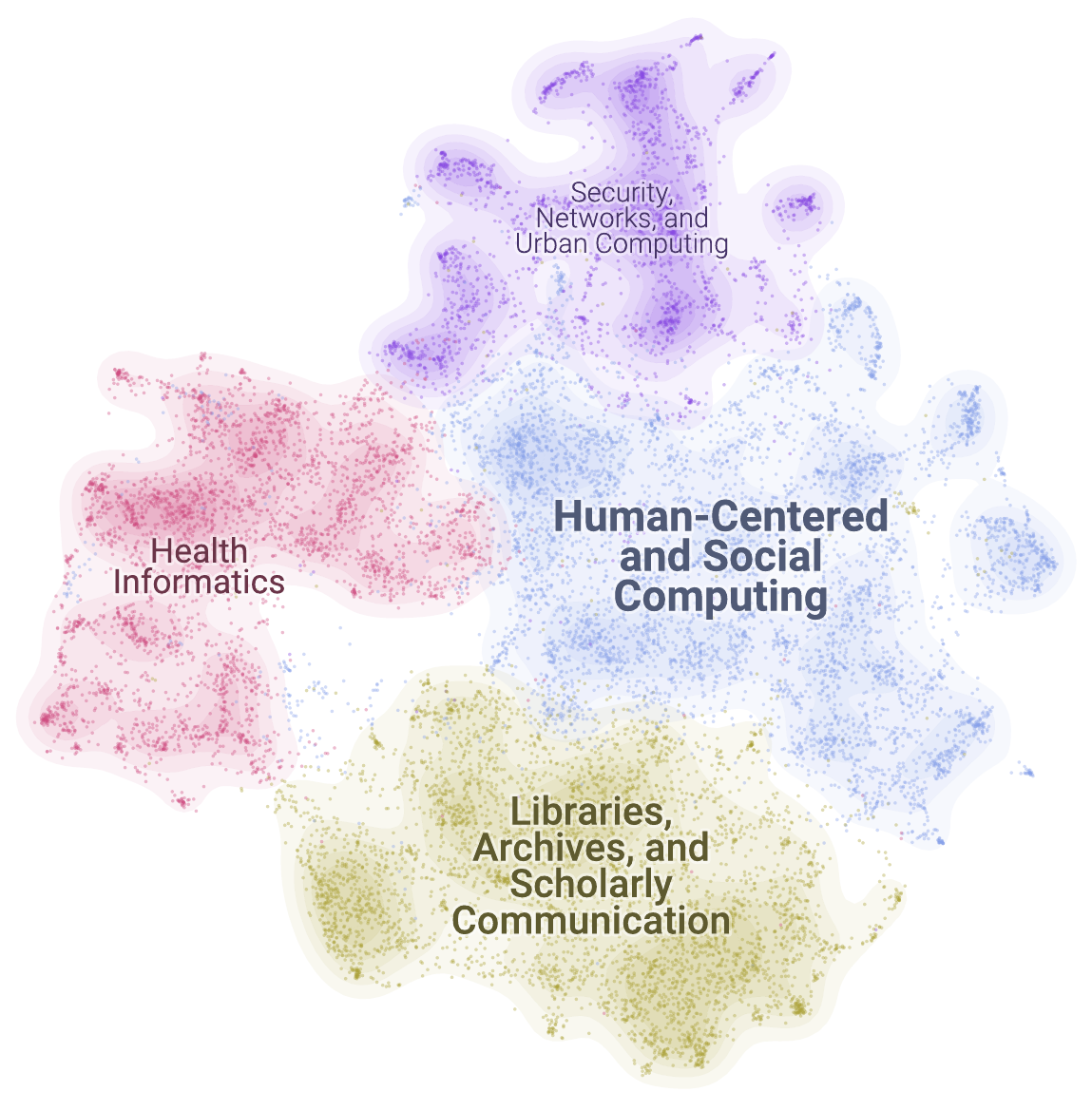}
    \caption{The research landscape at the coarse level ($K=4$). Publications and their positions are those of Figure~3 in the main text. Each area takes the color of its largest theme; the area labels were chosen by the authors (Section~\ref{sec:klevels}).}
    \label{fig:map_k4}
\end{figure}

\begin{figure}[!htbp]
    \centering
    \includegraphics[width=\linewidth]{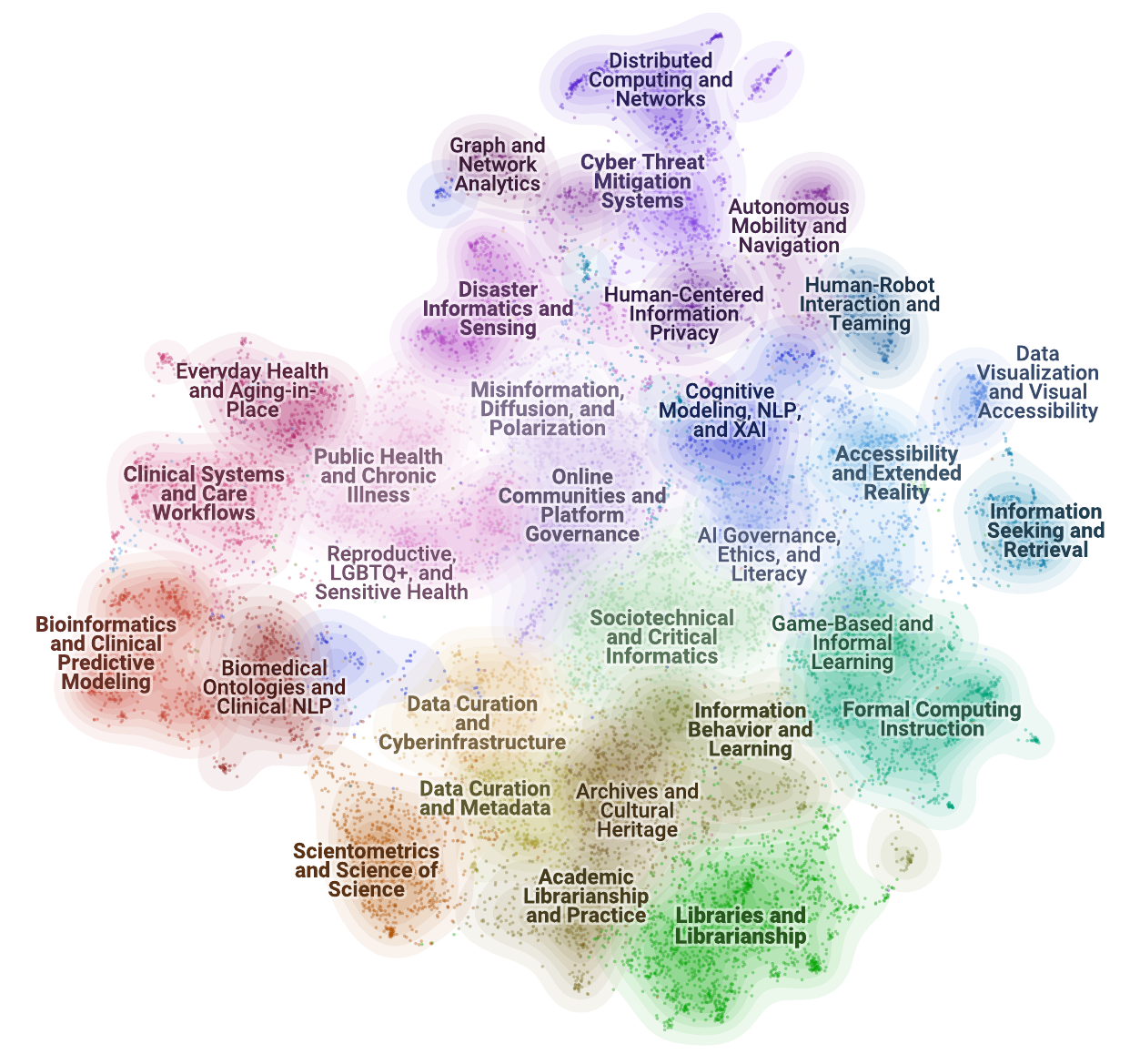}
    \caption{The research landscape at the fine level ($K=30$). Publications and their positions are those of Figure~3 in the main text. The parts of a split theme take shades of that theme's color; their labels were generated by Gemini 3.8 Flash.}
    \label{fig:map_k30}
\end{figure}

Because the measures support a range of solutions rather than 15 alone, we also show the themes at one coarser and one finer level. Both are cuts of the same Ward linkage as the 15 themes, so the three levels nest: each $K=4$ area is a union of themes, and each theme is a union of $K=30$ themes. Table~4 in the main text shows the four areas with their themes, Table~\ref{tab:k30} the finer themes of each theme, and Figures~\ref{fig:map_k4} and~\ref{fig:map_k30} draw the coarse and fine levels on the research landscape of Figure~3 in the main text.

To choose the two levels, we extended the recovery analysis of Section~\ref{sec:k14replication} to every cut from $K=2$ to $K=45$: for each cut, we counted the themes recovered in at least half of the 20 half models and the share of publications that these themes hold (Figure~\ref{fig:ksweep}e, which shows the cuts from $K=3$ to $K=40$), and we recorded the size of the smallest theme (Figure~\ref{fig:ksweep}f).

For the coarser level we chose $K=4$. No cut finer than $K=2$ has all of its areas recovered in at least half of the half models, and among all cuts with more than two areas, $K=4$ holds the largest share of publications in recovered areas (87\%; Figure~\ref{fig:ksweep}e). Libraries, Archives, and Scholarly Communication is recovered in all 20 half models, Health Informatics in 70\%, and Human-Centered and Social Computing in 60\%, but Security, Networks, and Urban Computing in only 20\%. At $K=3$, Health Informatics is recovered in 35\% of the half models. At $K=5$, the area that the cut adds, Social Media Information Dynamics together with Sociotechnical and Critical Informatics (2{,}313 publications), is recovered by no half model. We use this level to describe the broad areas of research, not as an alternative theme model: it scores low on coherence (0.084) and split-half stability (0.378, SD 0.090; Table~\ref{tab:ksweep}).
For the finer level we chose $K=30$. Its coherence (0.169) is higher than that of any coarser cut, and half of its publications (51\%) lie in themes recovered in at least half of the half models. It contains no small fragment: the smallest theme of every cut up to $K=38$ has at least 143 publications, and only at $K=39$ does Ward separate a group of 80 publications on blockchain, a third of them attributed to one faculty member (Figure~\ref{fig:ksweep}f). Its split-half stability (0.400, SD 0.022) is lower than that of the cuts from $K=8$ to $K=28$ (0.419 or more), as is that of every finer cut. Its smallest theme, Graph and Network Analytics, has 161 publications (1.0\%). Of the 30 themes, 8 are recovered in at least 70\% of the half models and 8 in 10\% or fewer.

A fine theme whose publications are exactly those of one of the 15 themes keeps that theme's label. We labeled the others, the 4 areas at $K=4$ and 24 themes at $K=30$, with Gemini 3.8 Flash (\texttt{gemini-3.8-flash}), adapting the prompt used for the 15 themes and giving the model every title in the area or theme. Each area received one label covering its themes, which we asked to differ from the labels of those themes. The parts of each split theme were labeled together, in one prompt, so that their labels distinguish them from one another. We kept the model's labels of the fine themes. We renamed the four areas, whose model labels were vague (the model's labels are given in parentheses): Libraries, Archives, and Scholarly Communication (Information Infrastructures and Stewardship), Human-Centered and Social Computing (Socio-Technical Information Interaction), Health Informatics (Integrative Health Information Science), and Security, Networks, and Urban Computing (Resilient and Trustworthy Urban Informatics).

At $K=4$, each area joins two to six themes (Table~4 in the main text). At $K=30$, nine themes split into two to four parts each, and six remain whole (Table~\ref{tab:k30}). Several splits separate recognizable subfields, such as Archives and Cultural Heritage within Archives, Metadata, and Data Sharing, and Human-Robot Interaction and Teaming within Human-AI Interaction and Visualization.

\begin{table}[!htbp]
    \centering
    \caption{The 15 Themes and Their Finer Themes}
    \label{tab:k30}
    \singlespacing
    \footnotesize
    \setlength{\tabcolsep}{4pt}
    \newcommand{\finer}[1]{\begin{tabular}[t]{@{}>{\raggedright\arraybackslash}p{7.3cm}r@{}}#1\end{tabular}}
    \begin{tabular}{>{\raggedright\arraybackslash\leftskip=1em}p{6.2cm}rl}
    \toprule
    \multicolumn{2}{l}{\textbf{$K=15$}} & \textbf{$K=30$} \\
    \cmidrule(lr){1-2}\cmidrule(lr){3-3}
    \multicolumn{1}{l}{\textbf{Theme}} & \textbf{$n$} & \finer{\textbf{Finer theme} & \textbf{$n$}} \\
    \midrule
    \multicolumn{3}{l}{\textit{Libraries, Archives, and Scholarly Communication} (4{,}647)} \\
    Archives, Metadata, and Data Sharing & 1{,}893 & \finer{Data Curation and Metadata & 515 \\ Academic Librarianship and Practice & 506 \\ Information Behavior and Learning & 469 \\ Archives and Cultural Heritage & 403} \\
    \addlinespace[2pt]
    Libraries and Librarianship & 1{,}514 & \finer{\textit{unchanged} & } \\
    \addlinespace[2pt]
    Scientometrics and Science of Science & 774 & \finer{\textit{unchanged} & } \\
    \addlinespace[2pt]
    Data Curation and Cyberinfrastructure & 466 & \finer{\textit{unchanged} & } \\
    \addlinespace
    \multicolumn{3}{l}{\textit{Human-Centered and Social Computing} (5{,}793)} \\
    Human-AI Interaction and Visualization & 1{,}330 & \finer{AI Governance, Ethics, and Literacy & 403 \\ Data Visualization and Visual Accessibility & 373 \\ Cognitive Modeling, NLP, and XAI & 370 \\ Human-Robot Interaction and Teaming & 184} \\
    \addlinespace[2pt]
    Social Media Information Dynamics & 1{,}283 & \finer{Misinformation, Diffusion, and Polarization & 788 \\ Online Communities and Platform Governance & 495} \\
    \addlinespace[2pt]
    Computing Education and Learning Technologies & 1{,}032 & \finer{Formal Computing Instruction & 662 \\ Game-Based and Informal Learning & 370} \\
    \addlinespace[2pt]
    Sociotechnical and Critical Informatics & 1{,}030 & \finer{\textit{unchanged} & } \\
    \addlinespace[2pt]
    Accessibility and Extended Reality & 583 & \finer{\textit{unchanged} & } \\
    \addlinespace[2pt]
    Information Seeking and Retrieval & 535 & \finer{\textit{unchanged} & } \\
    \addlinespace
    \multicolumn{3}{l}{\textit{Health Informatics} (3{,}362)} \\
    Human-Centered Health Informatics & 1{,}322 & \finer{Clinical Systems and Care Workflows & 865 \\ Everyday Health and Aging-in-Place & 457} \\
    \addlinespace[2pt]
    Digital Health Information Behavior & 1{,}032 & \finer{Public Health and Chronic Illness & 680 \\ Reproductive, LGBTQ+, and Sensitive Health & 352} \\
    \addlinespace[2pt]
    Biomedical Informatics & 1{,}008 & \finer{Bioinformatics and Clinical Predictive Modeling & 578 \\ Biomedical Ontologies and Clinical NLP & 430} \\
    \addlinespace
    \multicolumn{3}{l}{\textit{Security, Networks, and Urban Computing} (2{,}132)} \\
    Cybersecurity and Information Privacy & 1{,}121 & \finer{Cyber Threat Mitigation Systems & 505 \\ Distributed Computing and Networks & 366 \\ Human-Centered Information Privacy & 250} \\
    \addlinespace[2pt]
    Crisis Informatics and Urban Computing & 1{,}011 & \finer{Disaster Informatics and Sensing & 516 \\ Autonomous Mobility and Navigation & 334 \\ Graph and Network Analytics & 161} \\
    \bottomrule
    \end{tabular}

    \vspace{4pt}
    \begin{minipage}{0.95\linewidth}
    \footnotesize\textit{Note.} Italic rows are the four areas ($K=4$) of Table~4 in the main text, with their number of publications. The three levels are cuts of the same Ward linkage, so each theme is a union of finer themes. \textit{Unchanged} marks a theme whose publications form a single finer theme, which keeps the theme's label; the labels of the other finer themes were generated by Gemini 3.8 Flash.
    \end{minipage}
\end{table}

\section{Classification of Publication Venues}
\label{sec:venuefields}
A venue is the journal, book series, or conference series in which a publication appeared, with the editions of each conference merged into one venue; 16{,}713 of the 18{,}048 publications have one.
For journals and book series indexed in the Web of Science, we used their Web of Science subject categories; for conferences and other venues, the Fields of Research (FoR) that Dimensions assigns to their publications.
We mapped each Web of Science category and each FoR group to one of the ten venue fields by hand, following the OECD Fields of Science and Technology \parencite{oecd2015frascati,kaliuzhna2024mapping}.
Each venue then took the field of its categories or, when these map to more than one field, the field of most of its publications by FoR.
On checking the result, we corrected the field of three venues.
Of the 18{,}048 publications, 1{,}360 (7.5\%) have no venue field: 1{,}335, mostly books and chapters outside a series, have no venue, and 25 appeared in 19 venues with neither Web of Science categories nor Fields of Research.

\section{Robustness Checks}
\label{sec:robustness}
\textit{Schools as units.} The PERMANOVAs of the main text compare individual faculty members.
With each school as one unit instead, described by the mean of its faculty members' square-root profiles, the school types also differ in the research themes (pseudo-$F = 2.12$, $R^2 = .179$, $p = .024$) and in the venue fields (pseudo-$F = 3.72$, $R^2 = .276$, $p = .002$).

\textit{Research diversity without coverage adjustment.} Without first giving large and small schools the same chance to reveal their themes, the effective numbers of themes of the school profiles give medians in the same order as at equal sample coverage: \textsc{Computer} 10.4, \textsc{Information} 9.3, \textsc{Communication} 7.2, \textsc{Art\&Science} 6.2, and \textsc{Education} 5.6 (Kruskal--Wallis $H = 10.3$, $p = .036$).

\textit{Research diversity over time at equal numbers of faculty members.} The diversity of a school in a period rises with the number of its faculty profiles.
To compare schools and periods at equal size, we also computed it for five faculty members drawn at random without replacement from the school's faculty profiles of the period, as the effective number of the 30 finer themes in the mean of their profiles, averaged over 500 draws; the 37 of 219 school-periods with fewer than five faculty profiles are left out.
This weakens the association of diversity with the number of faculty profiles, from a Spearman correlation of 0.85 to 0.62 across the 182 school-periods with a value and from 0.55 to 0.16 within schools.
At equal numbers of faculty members, the diversity of the average school rose from 6.1 themes in 2013--2015 to 6.8 in 2025--2026, by 12\% ($+0.61$ per decade, 95\% interval $[0.07, 1.15]$), against 24\% in the full period profiles (from 8.2 to 10.2 themes).
No type's interval excludes zero; the slopes range from $+0.22$ in \textsc{Art\&Science} to $+0.71$ in \textsc{Computer} schools.

\clearpage
\begin{center}\textbf{References}\end{center}
\printbibliography[heading=none]